\documentclass[twocolumn,           % Format : preprint, twocolumn
               showpacs,            % Pacs : showpacs, noshowpacs
               preprintnumbers,     % Preprint: preprintnumbers,
               aps,                 % Society: ...
               prd,          	    % Journal Style : pra, prb, prc, prd, pre,
               a4paper,             % Size : a4paper, ...
               superscriptaddress,  % Affiliation (Title) : groupedaddress,
               unsortedaddress,     %                       superscriptaddress,
               nofootinbib,         % Footnote: footinbib, nofootinbib
               tightenlines,        % Remove additional spaces in a line
               floats               % Floating pictures and tables
               ]{revtex4}

\usepackage{graphicx}
\usepackage{latexsym}
\usepackage{amsmath,amssymb}        % amssymb includes amsfonts
\usepackage[draft=false]{hyperref}

\let\include\input

\begin{document}

%%--- DRAFTCOPY --------------------------------
%% Prints a large "DRAFT" diagonally across each page
%% Does not show up in TeXview
%% \typeout{Prints "DRAFT" on each page; does not show in TeXView}
%% \special{!userdict begin /bop-hook{gsave 200 30 translate
%% 65 rotate /Times-Roman findfont 216 scalefont setfont
%% 0 0 moveto 0.90 setgray (DRAFT) show grestore}def end}
%%------------------------------------------------

%======================================%
%<<<<<<<<<<<< TITLE PAGE >>>>>>>>>>>>>>%
%======================================%

\title{Primordial black holes in braneworld cosmologies: astrophysical
constraints}
\author{Dominic Clancy}
\affiliation{Astronomy Centre, University of Sussex,
             Brighton BN1 9QJ, United Kingdom}
\affiliation{HEP Theory Division, Physics Department, P.O. Box 2208, 
	University of Crete, Heraklion, GR-71003, Greece (present address)}
\author{Raf Guedens}
\affiliation{DAMTP, Centre for Mathematical Sciences,
 Cambridge University, Wilberforce Road,
            Cambridge CB3 0WA, United Kingdom}
\author{Andrew R.~Liddle}
\affiliation{Astronomy Centre, University of Sussex,
             Brighton BN1 9QJ, United Kingdom}
\date{\today}
\pacs{98.80.Cq \hfill astro-ph/0301568}
\preprint{astro-ph/0301568}

%======================================%
%<<<<<<<<<<<<< ABSTRACT >>>>>>>>>>>>>>>%
%======================================%

\begin{abstract}
In two recent papers we explored the modifications to
primordial black hole physics when one moves to the simplest braneworld model,
Randall--Sundrum type~II. Both the evaporation law and the cosmological
evolution of the population can be modified, and additionally accretion of 
energy from the background can be dominant over evaporation at high energies. In 
this paper we present a detailed study of how this
impacts upon various astrophysical constraints, analyzing constraints from the 
present
density, from the present high-energy photon background radiation, from 
distortion of the microwave background spectrum, and from
processes affecting 
light element abundances both during and after nucleosynthesis. Typically, the 
constraints on the formation rate of primordial black holes weaken as compared 
to the standard cosmology if black hole accretion is unimportant at high 
energies, but can be strengthened in the case of efficient accretion.
\end{abstract}

\maketitle
\newcommand{\lra}{\Leftrightarrow}
\newcommand{\Mnodim}{\left(\frac{M}{M_4}\right)}
\newcommand{\TBHnodim}{\left(\frac{T_{{\rm BH}}}{T_4}\right)}
\newcommand{\Tnodim}{\left(\frac{T}{T_4}\right)}
\newcommand{\lnodim}{\left(\frac{l}{l_4}\right)}
\newcommand{\lminnodim}{\left(\frac{l_{{\rm min}}}{l_4}\right)}
\newcommand{\lmaxnodim}{\left(\frac{l_{{\rm max}}}{l_4}\right)}
\newcommand{\tLnodim}{\left(\frac{t_{{\rm evap}}}{t_4}\right)}
\newcommand{\tFnodim}{\left(\frac{t_{{\rm i}}}{t_4}\right)}
\newcommand{\teqnodim}{\left(\frac{t_{{\rm eq}}}{t_4}\right)}
\newcommand{\gcosm}{g_{{\rm cosm}}}
\newcommand{\tevap}{t_{{\rm evap}}}
\newcommand{\TBH}{T_{{\rm BH}}}
\newcommand{\be}{\begin{equation}}
\newcommand{\ee}{\end{equation}}
\newcommand{\bea}{\begin{eqnarray}}
\newcommand{\eea}{\end{eqnarray}}
\renewcommand{\L}{\left(}
\newcommand{\R}{\right)}
\newcommand{\alphaevap}{\alpha_{{\rm evap}}}

%%%%%%%%%%%%%%%%%%%%%%%%%%%%%%%%%%%%%%%%%%%%%%%%%%%%%%%%%%%%%%%%%%%%%%%%%%%%%%%%
%%%%%%%%%%%%%%%%%%%%%%%%%%%%%%%%%%%%%%%%%%%%%%%%%%%%%%%%%%%%%%%%%%%%%%%%%%%%%%%%

\section{Introduction}

The idea that our observable Universe may be a brane embedded in a
higher-dimensional bulk is one which has deep ramifications for cosmology, and
which in particular may rewrite many of our ideas as to how the Universe evolved
during its earliest stages. In most cosmological contexts, the effects of the
braneworld scenario are restricted to early Universe phenomena, though they may
impact on present observations by modifying relics from the early Universe, such
as inflationary perturbations or the dark matter density. However, in a recent
paper (\cite{GCL}, hereafter referred to as Paper~I), we reported on an
exception to this rule; Primordial Black Holes (PBHs) formed during the early
Universe may still probe the bulk dimensions today. 

Our analysis was restricted to the simplest braneworld scenario,
Randall--Sundrum type II~\cite{RSII}, (henceforth RS-II), where there is a
single bulk dimension of anti-de Sitter form characterized by an AdS
radius of curvature $l$.
We showed that provided the AdS radius was sufficiently large, PBHs whose
lifetime was as long as the present age of the Universe could behave
as five-dimensional objects, and that this would lead to reductions
both in the mass and the temperature corresponding to a given lifetime.

In Paper I it was argued that at the time a density fluctuation
collapses to form a black hole, its mass will be of order the horizon
mass. Subsequently a braneworld black hole could undergo substantial
growth by accreting material from the cosmological background, as
shown by Majumdar \cite{Maj} and by Guedens et al (\cite{GCL2}, hereafter 
Paper~II). This accretion phase will typically
last until the standard cosmological regime is reached, after which the 
evolution is governed by Hawking evaporation. 

Once formed, PBHs will influence later cosmological epochs, leading to
a number of observational constraints on their allowed abundance.
These have been  extensively investigated in the case of the
standard cosmology~\cite{ZeldovichNovikov66}-\cite{GreenLiddle97} 
%\cite{ZeldovichNovikov66,Hawking71,CarrHawking74,Hawking75,Carr75,Carr76,
%Novikovetal79,PolnarevKhlopov80,CarrLidsey93,CarrGilbertLidsey94,
%GreenLiddle97} 
(see also Ref.~\cite{Carr85} for a review),
and the aim of the present paper is to reanalyze the main constraints
in the braneworld context. Because the temperatures of black holes evaporating
at a given epoch are modified, for the most part such constraints have to be
recomputed from first principles.

To outline the types of constraints that can arise, let us consider an epoch
labeled by cosmic time $t$. PBHs whose lifetime exceeds this time will
essentially still possess their initial mass and only contribute to the overall
energy density. As the observable Universe is close to flatness, a conservative
bound derives from the fact that the PBH mass density should not overdominate 
the
Universe.
PBHs with lifetimes of order $t$ are evaporating rapidly, producing bursts of
evaporation products. Limits can be obtained from imposing they should not
interfere disastrously with established processes such as those of
nucleosynthesis, or from the fact that these bursts have not been unambiguously
observed today \cite{MacGibbonCarr91}.
Even shorter-lived PBHs will have evaporated completely at an earlier stage.
If this happened well before the decoupling time of a particular species of
evaporation product, its Hawking radiation will thermalize with the
surroundings, boosting the photon-to-baryon ratio in the process \cite{ZelStar}.
In the case of evaporation after photon decoupling, the radiation spectrum
remains intact and subsequently redshifts. Thus constraints arise from the
cosmic
background radiation at high frequencies \cite{PageHawk, Carr76,
Fichtel, Chapline, MacGibbonCarr91}.
Finally, if the radiation is emitted in a certain time window before
photon decoupling, it cannot be fully thermalized and will distort the CMB
spectrum~\cite{Fixsen, SZ70, Nasel'skii}. If it is assumed black holes leave behind 
a stable relic, this can lead to different constraints~\cite{Bowick} but will not 
be pursued here.

At a given epoch, the impact that is
to be constrained is usually dominated by those PBHs with a lifetime
of order the cosmic time at the epoch in question. In the accretion phase soon 
after formation, their energy density will fall off more slowly than that of 
dust. 
After the accretion phase we can neglect the energy loss through evaporation 
until the last stages of their lifetime. Therefore, in the standard 
cosmological regime the PBHs will predominantly behave like dust. We conclude 
that the fraction of the total energy due to PBHs grows proportional 
to the scale factor in the radiation-dominated regime, whilst staying constant 
in the matter-dominated regime. Translating the observational constraint into 
an upper limit on the initial fraction in PBHs then typically gives extremely 
strong bounds. Furthermore, since the initial BH mass is of order the horizon 
mass, the limit on the initial PBH fraction in turn implies a limit on the 
amplitude of density fluctuations, on scales entering the horizon at the time 
of formation of the PBH.

In this paper we will reconsider constraints from the present density
of black holes with lifetimes exceeding the age of the Universe and
from the present photon background, as well as the constraint stemming from
the limits on the allowed distortion of the CMB spectrum. We will also
reconsider constraints arising from the effect of PBHs evaporating
during or after nucleosynthesis on the light element abundances.

%%%%%%%%%%%%%%%%%%%%%%%%%%%%%%%%%%%%%%%%%%%%%%%%%%%%%%%%%%%%%%%%%%%%%%%%%%%%%%%
%%%%%%%%%%%%%%%%%%%%%%%%%%%%%%%%%%%%%%%%%%%%%%%%%%%%%%%%%%%%%%%%%%%%%%%%%%%%%%%
% KEY EXPRESSIONS 
%%%%%%%%%%%%%%%%%%%%%%%%%%%%%%%%%%%%%%%%%%%%%%%%%%%%%%%%%%%%%%%%%%%%%%%%%%%%%%%
%%%%%%%%%%%%%%%%%%%%%%%%%%%%%%%%%%%%%%%%%%%%%%%%%%%%%%%%%%%%%%%%%%%%%%%%%%%%%%%

\section{The key expressions}

\label{sec:key}

We begin by reviewing the key results from Papers~I and II, which can
be consulted for the full details. Our Universe is taken to be a flat Friedmann 
brane, with the effective 4D
cosmological constant set to zero. The energy density will be
radiation dominated up to the time of matter domination in the more recent
past. Under these conditions, there is an early, high-energy regime in which
the scale factor, energy density, Hubble radius and horizon mass are
given in terms of cosmic time $t$ as
\begin{eqnarray}
a  =  a_{{\rm h}} \left(\frac{t}{t_{{\rm h}}}\right)^{1/4} \quad & ; & \quad 
\label{ahigh} \rho  =  \frac{3 M_4^2}{32 \pi \,t_{{\rm c}}\,t}\,,\label{rhoHE}\\
R_{{\rm H}}=4 t   \quad & ; & \quad M_{{\rm H}}=8 M_4^2\;
\frac{t^2}{t_{{\rm c}}}\,.\label{mhhigh} \nonumber
\end{eqnarray}
This is followed by a standard regime, in which
\begin{eqnarray}
a = a_{{\rm h}} \left(\frac{t}{t_{{\rm h}}^{1/2} t_{{\rm 
c}}^{1/2}}\right)^{1/2} \quad & ; & \quad \label{astand}
\rho  = \frac{3 M_4^2}{32 \pi\, t^2}\,,\\
R_{{\rm H}}=2 t  \quad & ; & \quad M_{{\rm H}}= M_4^2 t \,.\label{mhstand} 
\nonumber
\end{eqnarray}
Here, $t_{{\rm h}}$ is an arbitrary time in the high-energy regime and $t_{{\rm 
c}}=l/2$ is
the transition time between the regimes.

As said, we assume that PBHs form with masses approximately equal to
the horizon mass. To incorporate the uncertainty in the non-linear
process of black hole formation, we introduce a factor $f$ as
\begin{equation}
M_{{\rm i}}=f\; M_H(t_{{\rm i}})\,.\label{MM}
\end{equation}
In the following sections the constraints will be found to be  
quite insensitive to its precise value.
 
The main distinction to be made is whether the PBHs are effectively 4D
or 5D, which results from comparing the event horizon radius $r_0$
with the AdS radius $l$. Interestingly, with
$f \lesssim 1$ a PBH will be small ($r_0 \ll l$) if and only if it
formed in the high-energy regime. Similarly, it will be large and effectively 
4D throughout the bulk of its lifetime if and only if it formed in the standard
regime. The behaviour of such large black holes should reduce to that
of standard cosmology.\footnote{Recently, an interesting conjecture was made in 
which the radiation 
from RS-II black holes can be computed via an AdS/CFT equivalence
\cite{Tanaka, Emparanetal}. If
true, this would greatly increase the evaporation
rate for large black holes as there are many decay routes into conformal
field degrees of freedom, potentially leading to radically different PBH
phenomenology. However, their result remains a conjecture, and in this
paper we continue to adopt the traditional view of PBH evaporation as
described in our earlier papers \cite{GCL, GCL2}. It would be interesting to 
fully
analyze the modifications to constraints on PBH abundances using their 
evaporation law.}

Neglecting possible charges or rotation, the small PBHs are to
good approximation described as 5D Schwarzschild black holes, for which 

\begin{eqnarray}
r_0 & = &  \sqrt{\frac{8}{3 \pi}} \lnodim^{1/2} \Mnodim^{1/2} l_4 \,;
\label{bhrad} \\
\TBH & = & \frac{1}{2 \pi r_0}\,, \label{tbh} 
\end{eqnarray}
where an index 4 refers to Planckian quantities as measured on the
brane. The evolution in time is obtained from
\begin{equation} 
\frac{dM}{dt}=\left(\frac{dM}{dt}\right)_{{\rm acc}}+\left(\frac{dM}{dt}\right)_
{{\rm evap}} \label{acc+evap} \,,
\end{equation}
with the accretion and evaporation rates given by
\begin{equation} 
\left(\frac{dM}{dt}\right)_{{\rm acc}}= \frac{q}{2}\,\frac{M}{t} \,,
\end{equation}
and
\begin{equation} 
\left(\frac{dM}{dt}\right)_{{\rm evap}}=-\frac{\tilde g}{2}
\lnodim^{-1} \Mnodim^{-1} M_4^2 \,.\end{equation}

Following Paper~II, the factor $q \equiv 4F/\pi$ in the accretion term 
parametrizes the efficiency with
which the black hole accretes the cosmic radiation background, with $F=0$ 
corresponding to no accretion and $F = 1$ corresponding to perfect accretion of 
high-energy radiation from a uniform background. $F$ is therefore expected to 
lie between zero and one. 

In the
evaporation rate we have defined
\begin{equation}
\tilde g\approx 0.0024\;g_{{\rm brane}} +0.0012\;g_{{\rm bulk}} \,,
\end{equation}
where $g_{{\rm brane}}$ is the usual number of particle species, while
$g_{{\rm bulk}}= \mathcal{O}(1)$ is the number of bulk degrees of freedom
(in the simplest case just the five polarization states of the
graviton). The energy lost through Hawking evaporation is mainly
emitted onto the brane. As an example we mention the case where the black
hole emits only massless particles, for which $g_{\rm{{\rm brane}}}=7.25$ and
$\tilde g=0.023$.

Primordial black holes that are relevant for observational constraints
must have lifetimes greatly exceeding the cosmological transition
time $t_{{\rm c}}$. It is then an excellent approximation to neglect the
evaporation term in Eq.~(\ref{acc+evap}) until $t=t_{{\rm c}}$, and subsequently
to neglect the accretion term. The mass at the transition reads  
\be M(t_{{\rm c}}) \approx M_{{\rm i}} \L\frac{t_{{\rm c}}}{t_{{\rm 
i}}}\R^{q/2}.\label{mtc}\ee
We stress that this is the mass the PBH has at the effective onset of
evaporation. 
The total lifetime $\tevap$ is given by the 5D 
mass--lifetime relation  
\be \frac{\tevap}{t_4} \approx  \tilde g^{-1} \frac{l}{l_4}
\L\frac{M(t_{{\rm c}})}{M_4}\R^2\,.\label{tlife} \ee
By assumption 
$t_{{\rm evap}} \gg t_{{\rm i}},t_{{\rm c}}$. Therefore, for a PBH of a given 
lifetime this 
relation determines what the mass at the onset of 
evaporation should be, irrespective of the occurrence of accretion. Hence
including accretion will not change the expected temperature of PBHs 
evaporating at a given epoch. 

The 5D relations above are to be contrasted with the usual
4D results, in which accretion plays no significant part:
\bea
r_0 & = & \frac{2M}{M_4^2} \,; \\
\TBH(4{\rm D}) & = & \frac{M_4^2}{8\pi M} \,; \label{4Dtemp}\\
\frac{\tevap(4{\rm D})}{t_4} & \approx &
1.2\times 10^4\;g_{{\rm brane}}^{-1}\, \L\frac{M_{{\rm i}}}{M_4}\R^3 
\,.\label{4dlifetime}
\eea

For black holes of a given lifetime $t_{{\rm evap}}$, the question
arises if they are effectively 4D or 5D. They will be small (5D)
if the AdS radius is greater than a critical value, given by
\begin{equation}
l_{{\rm min}}(\tevap) =\tilde g^{1/3} \tLnodim^{1/3} l_4 \,.
\label{lminmax}
\end{equation}
For much smaller values of $l$, the standard 4D case is retrieved.
For example, all PBHs with lifetimes up to the present age of the
Universe would have been five dimensional throughout their evolution
provided that $l>10^{20}\;l_4$. As the experimental upper limit on the AdS
radius currently is quite weak ($l< l_{{\rm max}}\approx 10^{31}
l_4$~\cite{Hoyle}),
there is considerable room for 5D PBHs to play a role in the
cosmological history, including the present.

A simple reworking of the 5D relations expresses the BH mass and
temperature at the onset of evaporation in terms of $l$ and $t_{{\rm evap}}$:
\begin{eqnarray}
\frac{M(t_{{\rm c}})}{M_4} & = & \tilde g^{1/2}
\left(\frac{\tevap}{t_4}\right)^{1/2} \lnodim^{-1/2} \,,
\label{masslife}\\
\frac{\TBH}{T_4} & = & \sqrt{\frac{3}{32 \pi}}\; \tilde g^{-1/4}
\tLnodim^{-1/4} \lnodim^{-1/4}\,,\label{templife}
\end{eqnarray}

\begin{figure}[t]
\includegraphics[width=\linewidth]{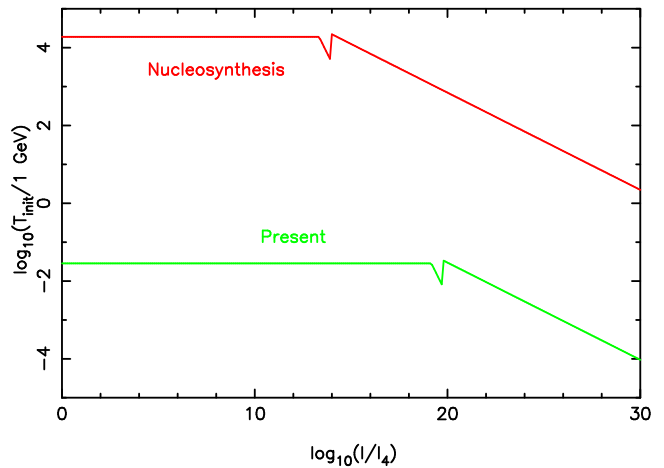}\\
\caption[inittemp]{\label{inittemp} The initial temperature of black holes
evaporating at the key epochs of nucleosynthesis ($t_{{\rm
evap}}\approx 100 \rm{s}$) and the present ($t_{{\rm
evap}}\approx 14$ Gyrs), shown as a function of $l$.}
\end{figure}

The temperature of PBHs evaporating today and at nucleosynthesis is
shown as a function of $l$ in Figure~\ref{inittemp}. Plots of the
mass are qualitatively the same. At the smallest
values of $l$, the temperature assumes the 4D
value, for $l\approx l_{{\rm min}}$ (which is equivalent to $r_0 \approx l$)
accurate description is uncertain through lack of exact solutions, while for
larger $l$ values the BH temperature is reduced. Since most of the
energy of a PBH is radiated at temperatures close to the start
temperature at the onset of evaporation, we conclude
that the evaporation products present at a certain epoch in cosmology
will be cooler if $l$ is sufficiently large. For example, PBHs
evaporating today would produce no massive particles, except in a high-energy
tail from the late stages of evaporation.

To conclude, we list the formation times $t_{{\rm i}}$ in terms of a given
lifetime $t_{{\rm evap}}$.
For a PBH that formed in the high-energy regime this is
\begin{eqnarray}
\L\frac{t_{{\rm i}}}{t_4}\R_{{\rm HE}} & = &
\frac{1}{4}f^{-1/2}\lnodim^{1/2} \left(\frac{M_{{\rm i}}}{M_4}\right)^{1/2}
\nonumber \\
 & = & 
\left[ 2^{q-8}\; \frac{\tilde g}{f^2} \lnodim^{1-q}
\left(\frac{\tevap}{t_4}\right)\right]^{1/(4-q)}\!\!. \;\;
\label{tF}
\end{eqnarray}
while in the standard regime one finds
\begin{equation}
\L\frac{t_{{\rm i}}}{t_4}\R_{{\rm ST}}=f^{-1} \frac{M_{{\rm i}}}{M_4}=0.04\; 
f^{-1}
g_{\mathrm{{\rm brane}}}^{1/3} \tLnodim^{1/3} \,.
\label{tFstand}
\end{equation}

It is worth bearing in mind that if black holes form after inflation, there is 
an upper limit on the mass scale coming from gravitational wave production. In 
Paper~I we showed this gives a lower limit on the black hole mass at formation 
of
\be M_{{\rm i}}>2\times 10^6 M_5.\label{lowlim}\ee
Using the lifetime--formation time relation Eq.~(\ref{tF}), Eq.~(\ref{lowlim}) 
implies a lower limit on the black hole's lifetime:
\be \frac{t_{{\rm evap}}}{t_4}
 >  \frac{f^2}{\tilde g} \, 2^{2+q/2} \times 10^{3 (4-q)}\,\lnodim^{(2q+1)/3}
\equiv \frac{t_{{\rm evap, min}}}{t_4}.\ee
For $l=10^{31} l_4$ and $100 \%$ efficiency this gives 
\be t_{{\rm evap}}>1100\,f^2\,s,\ee
i.e.~no black holes evaporating until after nucleosynthesis. But for an
efficiency lower than $86\,\%$ or when $l/l_4<10^{28}$ the lightest black
holes allowed by the lower mass limit can evaporate before $t=1s$.
Note however that the above limit may be a very conservative one. If inflation
was at an energy scale much lower than the allowed upper limit, the lightest 
permitted PBHs may evaporate at much later times still.

%%%%%%%%%%%%%%%%%%%%%%%%%%%%%%%%%%%%%%%%%%%%%%%%%%%%%%%%%%%%%%%%%%%%%%%%%%%%%%%
%%%%%%%%%%%%%%%%%%%%%%%%%%%%%%%%%%%%%%%%%%%%%%%%%%%%%%%%%%%%%%%%%%%%%%%%%%%%%%%

% CONSTRAINT FORMALISM

%%%%%%%%%%%%%%%%%%%%%%%%%%%%%%%%%%%%%%%%%%%%%%%%%%%%%%%%%%%%%%%%%%%%%%%%%%%%%%%
%%%%%%%%%%%%%%%%%%%%%%%%%%%%%%%%%%%%%%%%%%%%%%%%%%%%%%%%%%%%%%%%%%%%%%%%%%%%%%%

\section{Constraint formalism}
\label{sec:formalism}

Constraints on the allowed abundance of PBHs of a certain lifetime are 
formulated as upper bounds on their mass fraction. This mass fraction  
$\alpha_t(M_{{\rm i}})$ will be defined as
the ratio of the energy density due to PBHs of initial mass $M_{{\rm i}}$ and 
the
background radiation density, at a time $t \geq t_{{\rm i}}$:\footnote{The 
reader 
is warned that different notations are sometimes used, such as $\alpha=
\rho_{{\rm pbh}}/\rho_{{\rm tot}}$. This will not be important in the 
radiation-dominated phase, but the definitions do deviate in the 
matter-dominated phase.}
\be \alpha_t(M_{{\rm i}})\equiv \frac{\rho_{{\rm {{\rm pbh}},M_{{\rm 
i}}}}(t)}{\rho_{{\rm
rad}}(t)}.\ee
The initial and final mass fractions will be denoted $\alpha_{{\rm i}}$ and 
$\alpha_{{\rm evap}}$ respectively. 
The purpose of the following sections is to reconsider observational 
constraints on $\alphaevap$ at different cosmological epochs, trace them back 
to obtain constraints on the initial mass fraction $\alpha_{{\rm i}}$, and to 
compare
both types of constraint in the standard and braneworld cosmologies.
Let us consider a constraint imposed at a given epoch of cosmic time $\tau$:
\be \alpha_{\tau} < {L^{{\rm ST}}_{\tau}}\;\;\;\; {\rm or}\;\;\;\; \alpha_{\tau} 
<  
{L^{{\rm HE}}_{\tau}}, \label{evapconstraint}\ee
in the standard or braneworld scenario respectively.
Several constraints imposed at this stage do not depend on the individual
temperature or mass of the PBHs, but only on the total energy contained in
them or emitted by them, so that ${L^{{\rm ST}}_{\tau}}={L^{{\rm HE}}_{\tau}}$. 
Examples include the bound from the present mass density, from the 
distortion of the CMB, or the deuterium photo-disintegration constraint. An 
example of a bound that does depend on the 
individual characteristics of the PBH is the helium abundance constraint (see 
Section~\ref{sec:nuc}).   
 
To obtain the corresponding bound on the initial mass fraction requires
knowledge of the cosmic evolution. Note that
in the high-energy regime, i.e.~in the PBH accretion phase, we have  
\be 
\alpha_t \propto M(t)\, a(t) \,, 
\ee
whereas in the standard regime the mass fraction simply grows with the scale 
factor (until the final epoch of the PBH's lifetime). Tracing 
Eq.~(\ref{evapconstraint}) back to the time of formation, we get
\be 
\alpha_{i}<{L^{{\rm ST}}_{\tau}}\; \L \frac{a_{{\rm i}}}{a_{\tau}}\R_{{\rm
ST}}\equiv {L^{{\rm ST}}_{{\rm i}}}\,, \label{initconstraintst}
\ee
or
\be 
\alpha_{i}<{L^{{\rm HE}}_{\tau}}\;\; \frac{M_{{\rm i}}}{M(t_{{\rm c}})}\;
\L \frac{a_{{\rm i}}}{a_{\tau}}\R_{{\rm HE}}\equiv {L^{{\rm HE}}_{{\rm i}}} \,. 
\label{initconstrainthe}
\ee

\begin{figure}[t]
\includegraphics[width=\linewidth]{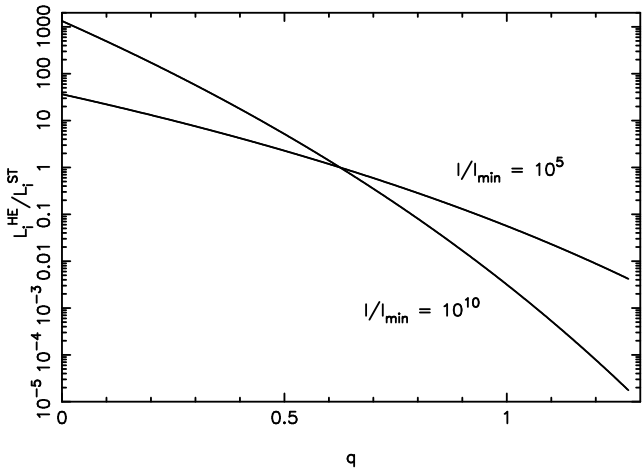}\\
\caption[limrat]{\label{limrat} The behaviour of Eq.~(\ref{compareinit}) as a 
function of the accretion parameter $q$ for two choices of the ratio $l/l_{{\rm 
min}}$, under the assumption that the imposed limit at evaporation is
identical in the high energy or standard treatment, i.e. 
$L^{{\rm HE}}_{{\rm evap}}=L^{{\rm ST}}_{{\rm evap}}$. This shows how 
constraints are modified compared to the standard 
cosmology in situations where the constraint at evaporation is unchanged. If 
accretion is negligible the constraints are weakened in the high-energy case, 
whereas for $q> 5/8$ accretion leads to the constraints strengthening in the 
braneworld case.}
\end{figure}

In order to compare the initial constraints Eqs.~(\ref{initconstraintst}) and
(\ref{initconstrainthe}), we express their ratio as
\be 
\frac{{L^{{\rm HE}}_{{\rm i}}}}{{L^{{\rm ST}}_{{\rm i}}}}= \frac{{\rm 
L^{{\rm HE}}_{\tau}}}
{{L^{{\rm ST}}_{\tau}}}\; \frac{M_{{\rm i}}}{M(t_{{\rm c}})}\; \frac{(a_{{\rm 
i}})_{{\rm HE}}}
{(a_{{\rm i}})_{{\rm ST}}}\,. 
\ee
The ratio of the initial scale factors can be expressed as
\be \frac{(a_{{\rm i}})_{{\rm HE}}}{(a_{{\rm i}})_{{\rm ST}}}=\frac{t_{{\rm 
i,HE}}^{1/4} \, t_{{\rm c}}^{1/4}}{t_{{\rm i,ST}}^{1/2}}.\ee
As mentioned in the introduction, the impact that is to be constrained at a 
certain epoch is usually dominated by PBHs with lifetimes of order the cosmic 
time at that epoch. We therefore take $\tau=t_{{\rm evap}}$. Using 
Eq.~(\ref{mtc}) for $M(t_{{\rm c}})$, and Eqs.~(\ref{tF}) or (\ref{tFstand}) 
for the lifetime--formation time relations, we obtain\footnote{We have omitted
powers of $f$, the ratio of the initial black hole mass and the horizon mass
(see Eq.~(\ref{MM})). Taking them into account we find 
${L^{{\rm HE}}_{{\rm i}}}\propto f^{-(1+2q)/2(4-q)}$ and 
${L^{{\rm ST}}_{{\rm i}}}\propto f^{-1/2}$. Since the accretion parameter $q$ is 
not 
expected to be larger than 1.5, these factors are of peripheral significance.} 
\be \frac{{L^{{\rm HE}}_{{\rm i}}}}{{L^{{\rm ST}}_{{\rm i}}}} \approx  
\frac{{L^{{\rm HE}}_{{\rm evap}}}}{{L^{{\rm ST}}_{{\rm evap}}}} \, 
\L\frac{l}{l_{\rm {min}}}\R^{(5-8q)/4(4-q)}. \label{compareinit}\ee
where $l_{{\rm min}}$ is the value of the AdS
radius at which quantities obtained in the HE scenario reduce to
the standard ones, see Eq.~(\ref{lminmax}). This equation is illustrated in 
Fig.~\ref{limrat}. Since $l>l_{{\rm min}}$, which 
constraint is the stronger is simply determined by the sign of $(5-8q)$, at 
least when ${L^{{\rm ST}}_{{\rm evap}}}={L^{{\rm HE}}_{{\rm evap}}}$. Thus the 
initial constraint will be stronger in the braneworld scenario if the accretion 
efficiency $F$ is more than $49 \%$, whilst becoming weaker for efficiencies
below $49 \%$. Whatever the accretion efficiency, maximum discrepancy with the
standard constraint is found when $l=l_{{\rm max}}\approx 10^{31} l_4$. Since
$l_{\rm min} \propto t_{{\rm evap}}^{1/3}$, this maximum discrepancy 
rises mildly when considering earlier epochs, i.e.~shorter-lived PBHs.

Note that if the calculations of Paper~II turn out to be invalid, and accretion 
is in fact not an important process even during the high-energy regime, the 
correct constraints are retrieved by filling in $q=0$ in the above formalism.

%%%%%%%%%%%%%%%%%%%%%%%%%%%%%%%%%%%%%%%%%%%%%%%%%%%%%%%%%%%%%%%%%%%%%%%%%%%%%%%%
%%%%%%%%%%%%%%%%%%%%%%%%%%%%%%%%%%%%%%%%%%%%%%%%%%%%%%%%%%%%%%%%%%%%%%%%%%%%%%%%

% OVERDOMINATION

%%%%%%%%%%%%%%%%%%%%%%%%%%%%%%%%%%%%%%%%%%%%%%%%%%%%%%%%%%%%%%%%%%%%%%%%%%%%%%%%
%%%%%%%%%%%%%%%%%%%%%%%%%%%%%%%%%%%%%%%%%%%%%%%%%%%%%%%%%%%%%%%%%%%%%%%%%%%%%%%%

\section{PBHs must not overdominate the Universe}
\label{sec:overclose}

For a particular value of $l$, Eq.~(\ref{tF}) or (\ref{tFstand}) with $t_{{\rm 
evap}}=t_0\,=\, 8 \times 10^{60}\, t_4$ gives the formation
time of PBHs that are evaporating today. Black holes formed
later are essentially still intact. Their density is constrained by the observed 
matter density in the present Universe; specifically, we are tracking the 
relative densities of PBHs to radiation, and we must ensure that, given the 
observed radiation density, this ratio does not imply that the PBH density 
exceeds the observed matter density of about 0.3 of the critical density. 
Phrased in this 
way, the constraint applies regardless of the presence of a cosmological 
constant, and indicates that for any PBHs surviving to the present we must have
\be 
\alpha_0(M)< \frac{0.3}{\Omega_{\gamma,0}} \,. \label{generalbound}
\ee 
The cosmic microwave background corresponds to a photon density as 
$\Omega_{\gamma,0} h^2=2.47 \times 10^{-5}$, with $h\approx 0.7$, and 
conservatively we can ignore the cosmic neutrinos.

{}From Eqs.~(\ref{initconstraintst}) and (\ref{initconstrainthe}) it is 
clear that a given observational constraint will give the most severe
initial constraint for the lightest PBHs it applies to, i.e.~those that formed
the earliest. Employing Eq.~(\ref{generalbound}) for PBHs that are about to 
evaporate today, $t_{{\rm evap}}\gtrsim t_0$, we find
\be 
\alphaevap < \frac{0.3}{\Omega_{\gamma,0}}\approx 6 \times 10^3 \equiv 
{L_{\rm evap}}\,. 
\ee
In the standard cosmology this is a bound on PBHs with mass $M_{{\rm i}} \approx 
2 \times 10^{19} M_4=4 \times 10^{14}$g, and the constraint  on the initial PBH 
mass fraction, Eq.~(\ref{initconstraintst}), reads~\cite{Carr75, Novikovetal79}
\begin{equation}
\alpha_{{\rm i}} < f^{-1/2}\times 10^{-18} \,.
\end{equation}

As examples in the braneworld scenario, we consider the case of maximum AdS 
radius, without accretion or with $100\%$ accretion respectively, i.e.~the
cases $l=l_{\rm max},\;q=0$, and $l=l_{\rm max},\;q=4/\pi$, and make use of 
Eq.~(\ref{compareinit}).
The first example gives $M_{{\rm i}}=10^{14} M_4=3 \times 10^9$g and
\be 
\alpha_{{\rm i}} < f^{-1/8} 10^{-14}\,,
\ee 
while the second example results in $M_{{\rm i}}=6 \times 10^5 M_4=10$g and
\be 
\alpha_{{\rm i}} < f^{-0.65} 10^{-23}\,.
\ee 
These are examples of the general trend discussed earlier, that going to the 
braneworld case without considering accretion weakens the constraint on the 
formation rate, but that including accretion can strengthen it, with fully 
efficient accretion leading to a more powerful constraint than in the
standard cosmology.

%%%%%%%%%%%%%%%%%%%%%%%%%%%%%%%%%%%%%%%%%%%%%%%%%%%%%%%%%%%%%%%%%%%%%%%%%
%									%
%**********************RADIATION SPECTRUM CONSTRAINT********************%
%									%
%%%%%%%%%%%%%%%%%%%%%%%%%%%%%%%%%%%%%%%%%%%%%%%%%%%%%%%%%%%%%%%%%%%%%%%%%

\section{The present photon spectrum}

If PBHs evaporate between the time of photon decoupling
($t_{{\rm dec}}\approx 10^{12}\,s$) and the present day, their radiation
spectra will not be appreciably influenced by the background
Universe, apart from being redshifted. Thus the spectra could
constitute a  fraction of the cosmic background radiation.
We will restrict attention to the photon
spectrum. In Refs.~\cite{MacGibbonWeber90, MacGibbonCarr91} it was pointed out 
that
black holes with temperatures above the QCD scale ($200-300$ MeV) should
emit quark and gluon jets that subsequently fragment into
particles whose rest mass is below the black hole temperature. This could
significantly alter the spectrum as compared to when the particles are
emitted directly as Hawking radiation. However, for large values of the AdS 
radius, the temperature of PBHs evaporating after decoupling
will never be above a few MeV, in which case the effect should be negligible.
We derive an expression for the spectral shape, and calculate it
explicitly in the assumption of a scale invariant initial PBH spectrum. We
then consider the constraint to be imposed on the peak of the
spectrum.

\subsection{The spectral shape}

The spectral photon number emitted onto the brane by a small black hole with
lifetime $\tevap$ is obtained from
\begin{equation}
\frac{dN}{dE}=\int \limits_0^{\tevap}
\frac{\sigma(E)}{2 \pi^2}\,\frac{E^2}{\exp\left[E/\TBH(t)\right]-1}\;dt
\,,
\end{equation}
with $\sigma(E)$ the emission cross-section for photons of frequency
$\omega=E$. In terms of the integration variable
\begin{equation} x\equiv \frac{E}{\TBH(t)} \,,
\end{equation}
and using the 5D relations of Section~\ref{sec:key}, this becomes
\begin{equation}\frac{dN}{dE}=\frac{9}{512\,\pi^4\,\tilde g} \lnodim^{-1}
M_4^3\, E^{-2} \int \limits_0^{x_i}
\sigma(x)\frac{x^3}{e^x- 1}\;dx\,,\label{xspec}\end{equation}
where $x_i=E/\TBH$ and $\TBH$ denotes the temperature of the black
hole at the onset of evaporation. In the high-frequency limit 
($E\gg\TBH$), all cross-sections reduce to the same value (see paper I)
\begin{equation}\sigma=4 \pi r_0^2=\frac{1}{\pi} E^{-2}
x^2.\end{equation}
In this limit, the spectral number becomes
\be \frac{dN}{dE}=\frac{\pi}{448}\,\tilde g^{-1}\lnodim^{-1} M_4^3\,E^{-4}.\ee
Note that the spectrum declines as $E^{-4}$, compared to
the $E^{-3}$ tail of the standard spectrum.

Now consider black holes
evaporating at a time $\tevap \geq t_{{\rm dec}}$. We will make the
approximation that all the energy gets released instantly, but will take
the spectrum into account. The black hole
mass fraction just before evaporation is given by
\begin{equation}\alpha_{{\rm evap}}=\alpha_{{\rm i}}\; \frac{M(t_{{\rm 
c}})}{M_{{\rm i}}}\;
\frac{a(t_{{\rm evap}})}{a(t_{{\rm i}})},\end{equation}
while the number density is
\begin{equation}n_{{\rm PBH}}(\tevap) = \alpha_{{\rm evap}}\;
\frac{\rho_{{\rm rad}}(\tevap)}{M(t_{{\rm c}})}.\end{equation}
The energy density in photons of energy $E$, emitted between
$\tevap$ and $\tevap+ dt_{{\rm evap}}$ is
\begin{equation}d\,U_{t_{{\rm ev}}}(E) \equiv n_{{\rm PBH}}(t_{{\rm
evap}})\;E^2\;\frac{dN}{dE}(\tevap)\;\frac{dt_{{\rm evap}}}{t_{{\rm evap}}}.
\end{equation}

We require the present total energy density in Hawking
photons at a certain energy scale $E_0$, denoted as $U_0(E_0)$. Radiation
emitted at $\tevap$ with energy $E$ will be redshifted to $E_0$ today provided
\begin{equation}E(E_0)=E_0\;\frac{a(t_0)}{a(\tevap)}.\end{equation}
Integrating over all times $t_{{\rm evap}}$ after decoupling , we have
\begin{eqnarray}U_0(E_0) & \equiv & \int
\limits_{t_{{\rm dec}}}^{t_0} d\,U_0(E_0)
\nonumber \\
& = &
\int \frac{a(\tevap)^4}{a(t_0)^4}\; d U_{t_{{\rm ev}}}\L E(E_0)\R .
\end{eqnarray}
We will assume $t_{{\rm i}}(t_0)<t_{{\rm c}}$, i.e.~all PBHs light enough to 
have evaporated
completely today were formed in the high-energy regime. Substituting
the relevant formulas then results in\footnote{Having omitted
powers of $f$ as discussed in previous sections.}
\begin{eqnarray} U_0(E_0) & = &
\frac{8^{1/4}}{6 \pi}\,\tilde g^{-9/16}
\left(\frac{t_0}{t_4}\right)^{-4/3}
\left(\frac{t_{{\rm eq}}}{t_4}\right)^{1/2}  \label{toten} \\
&& \hspace*{-36pt} \times \lnodim^{3/16} \; M_4^3 E_0^2 \nonumber \\
&& \hspace*{-36pt} \times \int \alpha_{{\rm i}}\;
 \L \frac{t_{{\rm evap}}}{t_4}\R^{-59/48} \frac{dN}{dE(E_0)}(t_{{\rm evap}})\;
Q (t_{{\rm evap}}) \; \frac{d\,t_{{\rm evap}}}{t_{{\rm 
evap}}}\,,\nonumber
\end{eqnarray}
with the factor $Q(t_{{\rm evap}})$ incorporating the effect of accretion. It 
is defined by
\be Q(t_{{\rm evap}})^{4 (4-q)/q} = 2^9 \tilde g^{-9/4} \lnodim^{27/4} 
\L \frac{t_{{\rm evap}}}{t_4}\R^{-9/4} \label{Q}\ee
and reduces to $Q=1$ if accretion is neglected.

The number spectrum of a black hole of initial temperature $T_{{\rm
BH}}$ peaks at an energy $E=b\,\TBH$, with $b\approx 5$ in the
standard treatment \cite{Page}. Therefore, unless $\alpha_{{\rm i}}$ is
sharply peaked at particular initial epochs, the main
contribution to the integral in Eq.~(\ref{toten}) is obtained when
$E(E_0)=b\, \TBH(\tevap)$, i.e.~from PBHs evaporating at 
$t_{{\rm evap}}=t_{{\rm main}}$, where
\begin{equation}t_{{\rm main}} \approx
\left(\frac{E_0}{b\;\TBH(t_0)}\right)^{12/5}\; t_0. \end{equation}
The contribution from PBHs evaporating earlier will come from the
high-frequency end of their spectrum, while PBHs evaporating at later
times will contribute radiation that originated in the low-frequency end.
Using the number spectrum Eq.~(\ref{xspec}) with $x_i=b$, we estimate
the total energy density at energy $E_0$ as
\begin{eqnarray} U_0(E_0) & \approx & 0.1\,\alpha_{{\rm i}} \, \tilde
g^{-7/10}b^{-69/20} \left(\frac{t_{{\rm eq}}}{t_4}\right)^{1/2}
\left(\frac{t_0}{t_4}\right)^{-17/10} \nonumber \\
& \;\; & \times \lnodim^{1/20} \left(\frac{E_0}{M_4}\right)^{29/20}\; 
Q(t_{{\rm main}}) \; M_4^4,\label{specmain}\end{eqnarray}
with
\bea 
Q(t_{{\rm main}})^{4 (4-q)/q} &  = & 2^9 \pi^{-27/10} b^{27/5} 
\tilde g^{-18/5}  \\
 & & \hspace*{-36pt} \times \lnodim^{27/5} \L\frac{t_0}{t_4}\R^{-18/5} 
\L\frac{E_0}{M_4}\R^{-27/5}\,.  \nonumber
\eea
The shape of the spectrum is seen to depend on the accretion parameter
$q$. Introducing $p$ through $U_0(E_0)\propto E_0^p$, the exponent
ranges from $p=1.45$ for $q=0$ to $p=0.82$ for $q=4/\pi$. For
comparison, in the standard scenario we have $p=1.5$ in this range of
the spectrum.\\
The time $t_{{\rm main}}$ as defined above will lie in the relevant time 
interval only for energies $E_0$ in the interval
\begin{equation}\left(\frac{t_{{\rm dec}}}{t_0}\right)^{5/12}\;b\;\TBH(t_0)<E_0
< b\;\TBH(t_0).\end{equation}
Radiation at lower frequencies will stem completely from
the low-frequency ends of the instantaneous spectra, with the dominant
contribution coming from PBHs evaporating around $t_{{\rm dec}}$. Its
intensity can generically be neglected as compared to the main frequency
range.\footnote{In addition, the Universe is still rather opaque at
these early times.} For energies $E_0>b\;\TBH(t_0)$, the dominant part comes
from the high-frequency tail of PBHs evaporating today. The number
spectrum Eq.~(\ref{xspec}) with $x_i=\infty$ is used to obtain
\begin{eqnarray} U_0(E_0) & \approx & 10^{-3}\,\alpha_{{\rm i}} \, \tilde
g^{-25/16} \left(\frac{t_{{\rm eq}}}{t_4}\right)^{1/2}
\left(\frac{t_0}{t_4}\right)^{-41/16} \nonumber \\
& \;\; & \times \lnodim^{-13/16} Q(t_0) \;\left(\frac{E_0}{M_4}\right)^{-2} 
M_4^4 \,.
\label{spechigh}\end{eqnarray}

\subsection{The observational constraint}

To make contact with observations, it is convenient to relate the integrated
energy density $U_0(E_0)$ to the spectral surface brightness $I(E_0)$ through
\begin{equation}I(E_0)=\frac{c}{4 \pi}\; \frac{U_0(E_0)}{E_0}\,.\end{equation}
The overall peak in the present spectrum is
at $E_{{\rm peak}}=b\,\TBH(t_0)$. Using $q=0,\;b=5,\;t_0=8\times 10^{60}\,t_4,\;
t_{{\rm eq}}=10^{-6}\,t_0 \;{\rm and}\; g=0.023$, we find for
$I(E_{{\rm peak}})$, expressed in the units ${\rm keV}\, {\rm cm}^{-2}\,
{\rm s}^{-1}\,{\rm sr}^{-1} \,{\rm keV}^{-1}$:
\begin{equation}I(E_{{\rm peak}})\approx 10^{23}\;\alpha_{{\rm 
i}}\;\lnodim^{-1/16}.
\end{equation}
The strongest constraint to be placed on the PBH spectrum from the
observed flux arises at  $E=E_{{\rm peak}}$. Employing the 5D
relations of Section~\ref{sec:key} for $t_{{\rm evap}}=t_0$, we find
that $E_{{\rm peak}}$ can range from a hundred MeV to
a few hundred keV, depending on the AdS radius. Thus if the AdS radius is
large, PBHs would mainly contribute to the hard X-ray background
\cite{FabianBarcons, Holt, Gruber}. The XRB exhibits a peak at around $30$
KeV, thought to be sourced by AGN \cite{AGN}, although the
issue is not settled at present. This peak arguably is at too low an
energy scale to be explained by braneworld PBHs, but note that
$T_{{\rm BH}}(t_0)\approx 50$ keV for $l=l_{{\rm max}}$ in our simple set-up.
There was long thought to be another excess of radiation at 1 to 10 MeV
(the so-called 'MeV bump')~\cite{Schonfelder,FabianBarcons}. However,
measurements made
with the COMPTEL telescope~\cite{Kappadath} have shown this to be due to an
instrumentation error, and the background is now believed to be smooth
over a wide range of energies. From the steepness
of the PBH spectrum Eqs.~(\ref{specmain}) and (\ref{spechigh}) we conclude 
that, if there is a spectrum of radiation due to evaporating PBHs, it should be
significantly fainter than the observed background.

As an example we consider $l=l_{{\rm max}}$, to give a peak energy
$E_{{\rm peak}}\approx 250$ KeV. The
observed surface brightness at this energy is given by~\cite{Gruber}
\begin{equation}I_{{\rm obs}}=0.1\; {\rm keV}\, {\rm cm}^{-2}\,
{\rm s}^{-1}\,{\rm sr}^{-1} \,{\rm keV}^{-1}.\end{equation}
The constraint $I(E_{{\rm peak}})<I_{{\rm obs}}$ then results in an upper
limit on the initial mass fraction:
\begin{equation}\alpha_{{\rm i}} < 10^{-23}.\end{equation}
For comparison, the corresponding constraint in the standard case is obtained
from the gamma-ray background at $E_{{\rm peak}} \approx 100 {\rm Mev}$ and reads
$\alpha_{{\rm i}} <10^{-27}$~\cite{PageHawk, Carr76, Fichtel, Chapline,
GreenLiddle97}.

An equivalent way to constrain the peak value of the photon 
spectrum is found by expressing the observed density of radiation of order a 
hundred keV in terms of its density parameter as 
\be \Omega \approx 10^{-9}.\ee
Assuming that the fraction of the PBH mass going into photons is roughly ten 
percent, one obtains
\be \Omega_{{\rm pbh}} < 10^{-8},\ee
which is identical in form to the constraints of Section~\ref{sec:overclose}, 
and simply strengthens them by 8  orders of magnitude.

%%%%%%%%%%%%%%%%%%%%%%%%%%%%%%%%%%%%%%%%%%%%%%%%%%%%%%%%%%%%%%%%%%%%%%%%%%%%%%%%
%%%%%%%%%%%%%%%%%%%%%%%%%%%%%%%%%%%%%%%%%%%%%%%%%%%%%%%%%%%%%%%%%%%%%%%%%%%%%%%%
% CMB SECTION
%%%%%%%%%%%%%%%%%%%%%%%%%%%%%%%%%%%%%%%%%%%%%%%%%%%%%%%%%%%%%%%%%%%%%%%%%%%%%%%%
%%%%%%%%%%%%%%%%%%%%%%%%%%%%%%%%%%%%%%%%%%%%%%%%%%%%%%%%%%%%%%%%%%%%%%%%%%%%%%%%

\section{Distortion of the Cosmic Microwave Background Spectrum}

Energy that is released at a time $t_{{\rm SZ}}\approx 10^{-10}\, t_0$
will fail to thermalize fully with the background radiation, modifying the
Planck law with a chemical potential $\mu$ \cite{SZ70,Fixsen}. 
The injected energy $\Delta E$ is related to $\mu$ as
\be \frac{\Delta E}{E}=0.71\; \mu, \label{inject}\ee
with $E$ the background energy. Observational results \cite{Fixsen}
suggest an upper limit on $\mu$ given by
\be \mu < 9\times 10^{-5}. \label{obscons}\ee
This can be used to constrain the fraction of PBHs
evaporating around the time $t_{{\rm SZ}}$, as will now be elaborated.
Since the period under consideration occurs well after neutrino decoupling, 
only the fraction $F$ of the PBH energy released onto the brane 
that does not go into neutrinos (or gravitons)
will be relevant for the above bound. For the largest values of the
AdS radius, the temperature of the black holes evaporating at $t_{{\rm
SZ}}$ is such that the electron is the only massive particle to be
produced, leading to an estimate for $F$ as
\be F\approx 0.5\,. \ee
Smaller values of the AdS radius presumably increase $F$ somewhat,
but this will not play a substantial part in our order of magnitude estimate
of the constraint. When the injected energy derives solely from
black hole evaporation, Eq.~(\ref{inject}) becomes
\be 
F\,\alphaevap(t_{{\rm SZ}})=0.71\, \mu \,. 
\ee

Translating the observational bound Eq.~(\ref{obscons}) into a bound on the mass
fraction of PBHs evaporating around the time $t_{{\rm SZ}}$
results in
\be \alpha_{{\rm evap}}< 1.3 \times 10^{-4}\equiv {L_{{\rm evap}}}\,. \ee
For the standard cosmology this gives an initial constraint
\be \alpha_{{\rm i}} < 10^{-21},\ee
as first obtained in~\cite{Nasel'skii}.
The corresponding initial limits in the braneworld scenario
for the example of $l=l_{{\rm max}}$, are ${L^{{\rm HE}}_{{\rm i}}}= 10^{-17}$ 
for $q=0$ and ${L^{{\rm HE}}_{{\rm i}}}= 10^{-28}$ for $q=4/\pi$.

%%%%%%%%%%%%%%%%%%%%%%%%%%%%%%%%%%%%%%%%%%%%%%%%%%%%%%%%%%%%%%%%%%%%%%%%%%%%%%

\section{Nucleosynthesis Constraints}

%%%%%%%%%%%%%%%%%%%%%GENERAL BACKGROUND STUFF%%%%%%%%%%%%%%%%%%%%%%%%%%%%%%%%%
In principle every cosmological era prior to the time of 
photon decoupling, at around $10^{12}$ seconds,                        
could have been affected by PBH particle
interactions~\cite{Carr76}. Of these, the era of 
standard big-bang primordial nucleosynthesis (SBBNS) is 
generally regarded as being the best understood and the most 
tightly constrained, and so presents the best place in which to look 
for such effects. To that end, a number of detailed investigations 
have already provided strong 
evidence to suggest that a range of nucleosynthesis reactions
and parameters should indeed have been modified in the presence of
evaporating PBHs, and that furthermore, for a 
sufficiently high density of PBHs, such modifications 
would ultimately have led to changes in the final light
element abundances 
\cite{VainerNaselskii77,Zeldovichetal77,Vaineretal78,MiyamaSato78,Lindley80,
RothmanMatzner81,KohriYokoyama00}. 
Broadly speaking, in this body of work existing observational 
limits on the light element abundances are used  
to put constraints on the size of such modifications, which then 
typically lead to strong constraints on the numbers of PBHs 
allowed to evaporate both during and after nucleosynthesis.
Here we re-examine two nucleosynthesis constraints in 
the context of the braneworld, 
namely the constraint on the increase in production 
of Helium-4 due to the injection of PBH hadrons~\cite{Zeldovichetal77}, 
and the constraint on the destruction of primordial 
deuterium by PBH photons~\cite{Lindley80}. Our main aim here is to get a 
reasonable estimate of how 
such constraints are modified, and not to perform a quantitatively 
precise calculation.

{}From Eq.~(\ref{4dlifetime}), one finds that
in the standard scenario, PBHs which evaporate during
nucleosynthesis, which we shall take to be between $\sim 1$ and
$400$ seconds, have initial masses which range from $\sim
10^9 \rm{g}$ at $1 \,\rm{s}$,  
to $\sim 10^{10} \rm{g} $ at $400\,\rm{s}$, with corresponding 
temperatures that range from $\sim 10^{4}\,\rm{GeV}$ to 
$\sim 10^{3}\,\rm{GeV}$. On the other hand, in the braneworld
picture we find from Eqs.~(\ref{masslife}) and (\ref{templife}) 
that PBHs which decay during nucleosynthesis have masses at 
the effective onset 
of evaporation in the range
\be
\tilde g^{1/2}\lnodim^{-1/2}
\lesssim \frac{M(t_{{\rm c}})}{10^{17}\;\rm{g}}\lesssim 
\;20 \;\tilde g^{1/2}\lnodim^{-1/2}
\label{nsmassrange}
\ee
with corresponding temperatures in the range
\be
5\, \tilde g^{-1/4} \lnodim^{-1/4}
\gtrsim \frac{T_{{\rm BH }}}{10^{7}\;\rm{GeV}}\gtrsim 
\tilde g^{-1/4} \lnodim^{-1/4}.
\label{nstemprange}
\ee
Strictly speaking, since $\tilde{g}$ is temperature dependent, 
the temperature of the evaporating braneworld PBH has to be solved iteratively.
However, since the temperature dependence of $\tilde{g}$ is in fact very 
weak over the range of interest, for most purposes an estimate of 
$\tilde{g}\sim 0.1$ usually suffices for all temperatures up 
to a few $\rm{TeV}$ or so.\footnote{For standard model fields on the 
brane and gravity in the bulk 
$g_{{\rm brane}}$ and $\tilde{g}$ respectively range from around 
$7.25$ and $0.023$, for $T\lesssim 1\,\rm{MeV}$ to
about $106.75$ and $0.26$ for $300 \rm{GeV}\lesssim T\lesssim 1\,\rm{TeV}$.
Therefore, over the temperature range, 
$1\,\rm{MeV} \lesssim T \lesssim 1\,\rm{TeV}$, 
we have $0.15\lesssim\tilde{g}^{1/2}\lesssim 0.51$ 
(factor of 4) and $2.9\gtrsim\tilde{g}^{-1/4}\gtrsim 1.4$ (factor of 2).
Hence, it is good enough in most calculations to take
$\tilde{g}\sim 0.1$.
In principle, however, we note that $\tilde{g}$ could take a
wider range of values at temperatures much higher than
the $\rm{TeV}$ scale.} Accordingly, for the maximum 
value of $l$ allowed by present observational
limits, $l \approx 10^{31} l_4$, the PBHs evaporating during
nucleosynthesis have masses which range from $\sim 10\,\rm{g}$
at 1 second to $\sim 200\,\rm{g}$ at 400 seconds and have
corresponding temperatures which range from around $1$ to $0.2$ GeV.
Thus, we may conclude that at the limit of the largest $l$
currently allowed, the `small' PBHs evaporating during nucleosynthesis
can evaporate a wide range of particles, e.g. massless particles,
neutrinos, electrons and positrons, muons, pions, etas, kaons, and
in addition, they are just hot enough to evaporate nucleons.\footnote{With 
regard to the evaporation of nucleons, 
it should be also noted that, in taking proper 
account of the grey body factors, it is expected that the actual 
Hawking temperature of these black holes will be hotter 
by a factor of a few than the pure black body 
temperatures quoted here.}
At the other extreme we can take the smallest value of $l$ for which 
Eqs.~(\ref{nsmassrange}) and (\ref{nstemprange}) are valid, i.e.~$l=l_{{\rm 
min}}$. Then the temperature and mass range of
the standard scenario are retrieved, as they must do.

%%%%%%%%%%%%%%%%%%%%%%%%%%%%%%%%%%%%%%%%%%%%%%%%%%%%%%%%%%%%%%%%%%%%%%%%%%%%
%									   %
%**********************HELIUM ABUNDANCE CONSTRAINT*************************%
%									   %
%%%%%%%%%%%%%%%%%%%%%%%%%%%%%%%%%%%%%%%%%%%%%%%%%%%%%%%%%%%%%%%%%%%%%%%%%%%%

\subsection{The Helium abundance constraint}

\label{sec:nuc}

The proton-to-neutron ratio, $n/p$, is a key parameter in
primordial nucleosynthesis (for reviews 
see~\cite{Bernsteinetal89,KolbTurner90,Pagel97,Sarkar96,SchrammTurner98,
Oliveetal00,Tytleretal00}) and small variations in its size
can have appreciable effects on the final values 
of the light element abundances, $^4\rm{He},\,^3\rm{He},
\,\rm{D},^7\rm{Li}$, that the theory predicts.

According to the standard picture, $n/p$ `froze out', i.e.~fell out of 
equilibrium and became fixed at a nearly constant value, 
at around $1$ second, when the weak interaction 
proton-neutron inter-conversion rate, 
$\Gamma_{\rm{n}\leftrightarrow \rm{p}}\sim G_F^2T^5$, fell below the Hubble 
expansion rate, 
$H\sim (G_N g_*)^{1/2} T^{2}$. Since $n/p$ had been kept near 
Boltzmann equilibrium up until this time, at freeze out it would have had 
a value of $\simeq \exp(-\Delta M_{{\rm np}}/ T_f)\simeq 1/6$, where 
$T_f \simeq \mbox{0.8 MeV}$ was the temperature at freeze-out and $\Delta 
M_{{\rm np}} \sim 1.293 \mbox{MeV}$ the neutron--proton 
mass difference. Subsequently the value of $n/p$ was then affected only by 
neutron beta-decay. In addition, however, neutrons and protons at this time 
were also undergoing collisions and were thereby able to form 
deuterium, via reactions such as $p+n \iff \rm{D}+ \gamma$. 
Initially, however, the energy and density of 
the photon background was sufficiently high so as to  
photodissociate all of the $\rm{D}$ that formed in this way.       
Only after the temperature of the Universe had fallen below 
about $0.8~\rm{KeV}$, which occurred at a time of around $100$ seconds, 
did the photodissociation rate drop below the 
$\rm{D}$ binding rate and nucleosynthesis start.
Once this so-called `D-bottleneck' had been breached, 
the binding of protons and neutrons into 
$\rm{D}$ was then quickly followed by the subsequent 
binding of $\rm{D}$ with protons and neutrons 
into tritium and $^3\rm{He}$ nuclei, and in turn the binding 
of these into $^4\rm{He}$. 
However, due to the absence of both 
stable mass-five and mass-eight nuclei, 
and in addition the existence of strong Coulomb-barriers to all 
reactions that could form nuclei with mass-six, seven, nine or heavier, 
these nucleosynthesis reactions were only able to proceed efficiently 
as far as $^4\rm{He}$. Consequently, the process essentially ended once 
all of the neutrons (i.e.~the fuel) had been bound into $^4\rm{He}$ nuclei,
leaving only a very small proportion bound in the form of $\rm{D}$ 
and $^3\rm{He}$, together with an even smaller proportion 
that were able to overcome the Coulomb barriers and go on 
further to bind into heavier nuclei, mainly $^7\rm{Li}$ and 
$^7\rm{Be}$. 

The production of $^3\rm{He}$ and $\rm{D}$ was so-called 
{\em rate limited}, i.e.~the precise quantities of these nuclei 
left over at the end of nucleosynthesis 
were determined by the efficiency of their binding into 
$^4\rm{He}$, or in other words by the reaction rates of 
the associated binding processes. Since the reaction rates 
were simply proportional to the values of the speed of light, 
the thermally-averaged cross-sections and 
the density of the baryons, i.e.~$\Gamma_a \sim c \langle \sigma, T \rangle 
n_a$, 
the efficiency of each process would 
have been sensitive to a variation in any of these quantities. 
Hence, in the absence of non-standard physics affecting the cross-sections, 
the determining factor would have been 
the baryon density, $n_b$.\footnote{It further follows that, as the 
baryon density was sensitive to the expansion 
rate of the Universe, any variation in the expansion rate would 
also have influenced the final distribution of 
the abundances.} Thus, the predictions of 
nucleosynthesis calculations are a function of essentially 
just one parameter, namely $n_b$, although this is usually  
expressed in terms of its photon number density normalized form, 
$\eta$. The greater the density of baryons, i.e.~the higher the value of 
$\eta$, the faster the reaction rates and the 
more efficient and complete the binding of neutrons into 
$^4\rm{He}$ and so the less left behind in the form of 
$^3\rm{He}$ and $\rm{D}$ and vice-versa.

The production of the heavier nuclei in nucleosynthesis are 
likewise sensitive to $\eta$, but display a more complex dependence 
on its value. 

By the time nucleosynthesis actually started
at $100$ seconds, $\beta$-decay had lowered $n/p$ 
to around $1/7$. Given that nucleosynthesis ended once virtually 
all of the neutrons had been bound into $^4\rm{He}$, 
it therefore follows that the {\em mass fraction} 
of $^4\rm{He}$ at the end of nucleosynthesis, 
$Y_{{\rm p}}= 4n_{^4\rm{He}}/n_{{\rm 
b}}$, should have been approximately twice the value of the 
neutron-to-baryon ratio when it began, namely 
\be
Y_{{\rm p}} \simeq \frac{2n}{n+p}=\frac{2\,n/p}{1+n/p}\simeq \mbox{0.25}.
\ee
Although this estimate may seem somewhat crude, it is in fact  
to first order in good agreement with a full 
numerical treatment and also with the current observational bounds.  
The production of $^4\rm{He}$ has only a mild sensitivity 
to $\eta$, since it is guaranteed that virtually all of 
the neutrons will end up in $^4\rm{He}$ whatever 
else happens. The small dependence that it does display 
derives from the fact that higher $\eta$ values allow 
earlier D-bottleneck breaching times, implying that 
nucleosynthesis can start sooner and therefore 
with higher initial values of $n/p$.    
Aside from the $\eta$ dependence, the actual 
uncertainty that currently arises in the {\em theoretical} 
prediction of $Y_{{\rm p}}$, is of order $0.2\%$ 
($\sigma_Y=0.0005$)~\cite{Fiorentinietal98}, 
and comes primarily from the (now small) uncertainty 
in the neutron lifetime~\cite{KolbTurner90}, 
which is presently estimated to 
be $\tau_{{\rm n}}=885.7\pm 0.8\,\rm{s}$~\cite{Casoetal98}. 

During the intervening period between the end of nucleosynthesis 
and the present, the universal light element abundances are 
{\em all} believed to have undergone a certain amount of chemical evolution 
due to the effects of stellar processing. 
In order to estimate the primordial abundances today, therefore, 
it is generally desirable to seek out astrophysical 
sites which have been the least affected by this. 
In particular, since metallicity is generally  
expected to be correlated with the degree of stellar 
processing that has taken place~\cite{Pagel92,Izotovetal99}, 
sites with low metallicity are 
thought to be good targets for observation.
It is then hoped that, with a sufficient understanding 
of the intrinsic physics of such sites, 
one can extrapolate to zero metallicity to obtain 
the primordial values. The best current observational estimates 
of the cosmic primordial $^4\rm{He}$ abundance are believed to 
come from studies of helium and hydrogen recombination lines 
in low metallicity clouds of ionized hydrogen, so-called {\em HII} regions, 
which reside in blue compact galaxies~\cite{HIIrefs}. 

Presently, however, estimates of the primordial 
$^4{\rm He}$ abundance, extrapolated from observations of these regions 
by the different groups of observers, exhibit central values 
that differ by significantly more than their quoted 
statistical errors. This situation is generally believed to be 
indicative of the fact that some or all of these estimates are dominated 
by systematics~\cite{HIIsystematics}. 
Though it remains possible that the scatter could also be evidence 
for a genuine variance in the primordial values themselves. 
In light of this general uncertainty concerning the value 
of $Y_{{\rm p}}$, we shall adopt here the compromise value, derived 
from re-analysis of the data by Olive et al.~\cite{Oliveetal00}, who 
proposed that   
\be
Y_{{\rm p}} = 0.238\pm 0.002\pm 0.005.
\ee
Here the first error is statistical while the second represents 
an estimate of the overall systematic uncertainty 
in modelling the physics of the HII regions. The errors 
are compatible with all the data reviewed. 
In the context of SBBNS, the above estimate of $Y_{{\rm p}}$ 
alone is then consistent with $1.2\leq \eta\leq 6.3$.\footnote{Taking account 
also 
of the current observational estimates of the 
other primordial abundances yields the smaller `concordance 
interval' of $2.6\leq\eta_{10}\leq 6.2$ \cite{Sarkar96}.}   

As first indicated by Zel'dovich et al.~\cite{Zeldovichetal77}
the situation just described could have been radically different 
if one allowed for the possibility of a population of PBHs, which were 
hot enough to evaporate nucleons during nucleosynthesis, 
i.e.~PBHs with
$T_{\rm{BH}}\gtrsim 2$GeV, since in such an eventuality it turns out 
(as we shall outline below) that $n/p$ would have continued to increase 
after the weak interaction freeze-out time, giving rise to the possibility of 
a significantly higher yield of $^4\rm{He}$. Based on observational data 
of the time Zel'dovich et al.~calculated that 
an increase in $n/p$ of more than about 80\% 
(from its presently accepted value), would have resulted 
in an unacceptable over-production of primordial 
$^4\rm{He}$.\footnote{Zel'dovich et al. took unacceptable to mean
$Y_{{\rm p}}\gtrsim 0.4$.} 
Consequently they were able to put a 
constraint on the mass fraction of PBHs evaporating at the time 
of nucleosynthesis and also to translate this into a constraint on 
the initial mass fraction.\footnote{In addition to studying the 
effects of PBH injected baryons on the value 
of n/p, Zel'dovich et al. also studied the 
effects of helium spallation by PBH baryons. This they argued 
would have lead to an increase of the deuterium abundance; a claim 
that was later supported by the numerical studies of 
Vainer et al.~\cite{Vaineretal78} and Rothman and 
Matzner~\cite{RothmanMatzner81}. 
However, we shall not discuss this effect here.} A numerical treatment of 
the effects suggested by Zel'dovich et al., which 
involved integrating the system of nucleosynthesis 
reaction equations modified by appropriate terms to account 
for the (annihilation and spallation) effects 
of injected PBH baryons, was later carried 
out by Rothman and Matzner~\cite{RothmanMatzner81}. The 
results of these simulations broadly concurred with 
the semi-analytical estimates of Zel'dovich et al. and  
provided both an improved quantitative and qualitative 
understanding of the PBH effects they had suggested.     
     
More recently, however, it has been argued that the majority of 
baryons emitted by PBHs would not be emitted directly 
as nucleons and mesons, as Zel'dovich et al. had assumed, but 
rather via the fragmentation of a QCD quark--gluon jet~\cite{MacGibbonWeber90}. 
An analysis of this mode of injection, which utilizes 
the formalism of Reno \& Seckel~\cite{RenoSeckel88}, has been carried out by 
Kohri \& Yokoyama~\cite{KohriYokoyama00}. 
Attempting to take proper account of QCD effects 
in this way, they find constraints on the mass-fraction in PBHs 
that are typically one or two orders of magnitude stronger.\footnote{We 
note that this analysis could in principle 
be easily extended to the present context. Moreover, 
in contrast to the case of standard PBHs,   
given the somewhat lower Hawking temperature of braneworld PBHs 
evaporating at nucleosynthesis, such an analysis would not entail 
a large extrapolation of the behaviour of QCD from presently 
observed experimental regimes.}
An alternative proposition, made by Heckler~\cite{Heckler97a,Heckler97b}, 
however, is that the QCD jets would not directly fragment into hadrons. 
Instead, the particles emitted by a PBH would form a dense plasma, 
through which the quarks and gluons would lose energy via QCD 
bremsstrahlung and pair production, and this in turn would give rise to a 
photosphere, which could then be constrained by observations.   
     
However, since we are here interested in only obtaining a 
first semi-analytic estimate of PBH effects in the braneworld case, 
we shall ignore such QCD effects and follow the 
original approach of Zel'dovich et al.\cite{Zeldovichetal77}, keeping in mind its drawbacks and 
limitations. 
In order to reconsider this constraint for braneworld primordial 
black holes, we shall start by estimating the number of particles
evaporated by a braneworld PBH during nucleosynthesis.
The total number density of emitted particles, $N_{{\rm em}}$,
resulting from the complete evaporation of a population of
PBHs of some given initial mass, may be expressed as\footnote{Note: In this
section we use a notation in which $N_{{\rm em}}$ is a number {\em density}
and we reserve $n$ specifically to mean the neutron number density.}
\be
N_{{\rm em}}=\frac{\rho_{{{\rm pbh}}}}{\left<E_{{\rm em}}\right>},
\label{nem}
\ee
where $\left<E_{{\rm em}}\right>$ is the average energy of the emitted
particles. The ratio of the energy density in PBHs
at evaporation to the background radiation energy density, $\rho_{{{\rm 
pbh}}}/\rho_{{\rm rad}}$,
is therefore \cite{Zeldovichetal77}
\be \alpha_{\rm{evap}}=
\frac{\left<E_{{\rm em}}\right>\, N_{{\rm em}}}{\left<E_{{\rm rad}}\right>\,
N_{{\rm rad}}},
\ee
where $\left<E_{{\rm rad}}\right>$ and $N_{{\rm rad}}$ are similarly
the average energy and number density of the particles comprising
the background cosmological radiation fluid. 

To good approximation, the ratio of the average energies above is given by
the ratio of the PBH temperature at the onset of evaporation
to the background temperature at evaporation. 
For the four-dimensional PBHs of standard cosmology, using 
Eqs.~(\ref{4Dtemp}) and (\ref{4dlifetime})
and applying the standard cosmological temperature-time 
relation 
\be
\frac{t}{t_4}= \left(\frac{45}{16\pi^3}\right)^{1/2}\,g_{\rm{cos}}^{-1/2}
\,\left(\frac{T}{T_4}\right)^{-2}\,,
\ee
(see e.g. \cite{KolbTurner90}) the total emitted number density 
at any time $t=t_{{\rm evap}}$ during nucleosynthesis can be expressed 
as \cite{Zeldovichetal77}
\be
N_{{\rm em}}\approx \alpha_{\rm{evap}}\times
0.13 \times \,g_{{\rm brane}}^{1/2}\, g_{{\rm cos}}^{-1/4}
\left(\frac{M_{{\rm i}}}{M_4}\right)^{-1/2}\,N_{{\rm rad}}.
\label{Nemitted}
\ee
In the braneworld scenario, using Eqs.~(\ref{tbh}) and (\ref{tlife})
results in
\be
N_{{\rm em}}\approx
\alpha_{\rm{evap}}\left(\frac{320}{\pi}\right)^{1/4}
\!
\tilde g^{1/2} g_{{\rm cos}}^{-1/4}\,
\left(\frac{M(t_{{\rm c}})}{M_4}\right)^{-1/2} N_{{\rm rad}}.
\label{nem2}
\ee
Evidently, Eqs.~(\ref{nem2}) and (\ref{Nemitted}) have
the same functional form. However, it is important to 
recall that both the mass and temperature of PBHs 
of a given lifetime are reduced in the braneworld
scenario. Taking the mass reduction in account, Eqs.~(\ref{Nemitted}) and
(\ref{nem2}) show that the emitted number density for a given mass fraction
$\alpha_{{\rm evap}}$ is increased in the braneworld scenario. But this
is obvious from the temperature reduction, which reduces the average
energy $\left<E_{{\rm em}}\right>$ per emitted particle.

Having found an estimate of the total number density of particles
evaporated, we now turn our attention to the interaction of these
particles with the cosmological background \cite{Zeldovichetal77}.
In the simplest case (i.e.~in absence of any PBH chemical potentials)
one expects PBHs to emit nucleons and anti-nucleons
with equal measure.
First let us consider the case of the neutrons and anti-neutrons.
An anti-neutron emitted by a PBH has two
possible fates; it may either annihilate with a background neutron, or
alternatively annihilate with a background proton. At high momenta the
cross-sections for these processes are essentially the same. The first
of these possibilities leads to no net change in the neutron-to-proton
ratio. This follows from the fact that on
average for every anti-neutron emitted by a PBH there is also a neutron
emitted, thus the background neutron which is annihilated will in effect
only be replaced on average by another neutron also emitted by the PBH.
On the other hand, by the same reasoning, in the second interaction the
background proton
is effectively replaced with a PBH neutron.
Moreover, because there are six times as many protons
in the cosmological background as neutrons, this latter reaction
is six times more likely than the former. Hence, on average we expect that
{\em for every
seven neutrons emitted, one simply replaces a background neutron and the
other six replace background protons}, thus increasing the neutron-to-proton
ratio.

The story for the protons and anti-protons emitted by the PBH
is similar. An anti-proton emitted by a PBH 
may either annihilate with a background proton, or with
a background neutron. Here again it is apparent
that the first case leads to no net change in the
neutron-to-proton ratio, as the annihilated background proton
will in effect just be replaced by a PBH proton, whereas 
in the second interaction, a background neutron is effectively replaced
with a PBH proton. In this case, however, the 
latter reaction is six times {\em less} likely than the former. Hence,
we expect that {\em for every seven protons emitted, six simply replace
background protons while only one replaces a background neutron.}

To summarize, on average every seven anti-neutrons and seven anti-protons
emitted by an evaporating PBH (along with equal numbers of their
anti-particles) effectively convert twelve background protons and two
background neutrons into seven protons and seven neutrons.
Hence, the change in the background neutron number density,
$n_{{\rm c}}$, is
\be
\delta n_{{\rm c}}=\frac{6}{7}\,n_{{\rm em}}-\frac{1}{7}\,p_{{\rm em}}
\approx \frac{5}{7}\,n_{{\rm em}},
\ee
where $n_{{\rm em}}$ and $p_{{\rm em}}$ are respectively the number 
densities of the neutrons and protons emitted by evaporating PBHs,
which to a good approximation will be the same.
Similarly, we find that $\delta p_{{\rm c}} \approx -({5}/{7}) n_{{\rm em}}$,
so that $\delta(n+p)=0$.

As discussed above, current observational estimates suggest that 
the $^4\rm{He}$ abundance, $Y_{{\rm p}}$ should be $23.8\pm 1.1$\%, 
where we have added the errors in quadrature and quoted 
the $2\sigma$ value. Assuming this to be a legitimate 
conservative estimate, then it obviously follows that the 
largest value that could at present be accommodated 
by a nucleosynthesis scenario that included the effects of 
PBH evaporations must be $24.9 \%$. 
The question that we wish to answer, however, 
is how much could PBH effects have actually contributed to $Y_{{\rm p}}$. 
To ascertain this we use an independent piece of information, namely 
the lower bound on $\eta$ from observations of 
Ly$\alpha$ absorption in quasar spectra~\cite{Weinbergetal97}, 
which states that $\eta_{10}\gtrsim 3.4$. If we take this bound 
at face value, then since SBBNS conserves $\eta$ and PBH 
evaporations decrease it, it follows therefore that 
nucleosynthesis could not have started at a lower value 
of $\eta$ than this. Now for this value of $\eta$ SBBNS 
predicts a value of $Y_{{\rm p}} \gtrsim 24 \%$. 
Moreover, since nucleosynthesis with PBHs will 
always produce a higher value of $Y_{{\rm p}}$, it is impossible 
to have a nucleosynthesis scenario 
which incorporates PBH effects for $\eta = 3.4$ 
such that $Y_{{\rm p}}$ will be less than this. Thus a necessary 
condition is that PBHs could not have increased the 
value of $Y_{{\rm p}}$ by more than about $0.9 \%$.        
Hence, we may estimate
that PBHs evaporating during nucleosynthesis may only increase
$Y_{{\rm p}}$ by as much as about 1\%, or
equivalently that
\be
\delta \left(\frac{2n}{n+p}\right)=\frac{2\delta n}{n+p}\approx \frac{5}{7}\,
\frac{2 n_{{\rm em}}}{n+p}< \frac{1}{100}.
\label{deltannb}
\ee
For all presently allowed values of $l$,
we find that PBHs which evaporate during nucleosynthesis emit nucleons.
In general, however, it is not known exactly what proportion of
the particles injected into the background by the PBH will
constitute nucleons. In the standard scenario, however,
Carr~\cite{Carr76} has estimated that around 20\% of the particles
emitted by a PBH decaying during nucleosynthesis will ultimately
go into nucleons and anti-nucleons.
Here, therefore, we shall assume that the PBHs
will emit a fraction $\mathcal{F}N_{em}$ of the total
emitted particles in nucleons and anti-nucleons,
and assume similarly that $\mathcal{F}\lesssim 0.2$. 
Therefore, since $n_{{\rm em}} \approx \mathcal{F}N_{\rm{em}}/4$, it
follows from Eq.~(\ref{deltannb}) that we must have
\be
\frac{N_{{\rm em}}}{n+p}\sim \frac{N_{{\rm em}}}{n_{{\rm 
b}}}<\frac{2.8}{100\mathcal{F}}.
\ee
Substituting for $N_{{\rm em}}$ using Eq.~(\ref{Nemitted}) gives the standard 
constraint \cite{Zeldovichetal77}
\be
\alpha_{\rm{evap}}
<\frac{0.22}{ \mathcal{F}}
\left(\frac{g_{\rm{cos}}}{g_{{\rm brane}}^2}\right)^{1/4}
\left(\frac{M_{{\rm i}}}{M_4}\right)^{1/2}\eta_{\rm{evap}},
\label{nem3}
\ee
where $\eta_{\rm{evap}}=n_{{\rm b}}/N_{{\rm rad}}$, is the baryon-to-photon 
ratio at evaporation. If one further assumes that $\eta$ is fixed from 
evaporation to the present day,\footnote{Strictly such an assumption 
is only applicable in absence of further PBH evaporations after 
nucleosynthesis, as these would contribute 
radiation to the background and so lower the baryon-to-photon 
ratio~\cite{ZeldovichStarobinskii76}. However, even if this were to occur, 
such a change is in any case constrained to be fairly small 
and so should  not affect our result much.} 
i.e.~$\eta_{{\rm evap}}=\eta_0$, this may also be written as
\be
\alpha_{\rm{evap}}< 6\times 10^{-9}\,
\frac{\Omega_{{\rm b}} h^2}{\mathcal{F}}
\,\left(\frac{g_{\rm{cos}}}{g_{{\rm brane}}^2}\right)^{1/4}
\left(\frac{M_{{\rm i}}}{M_4}\right)^{1/2},
\label{SCalphaeHeconstraint}
\ee
where we have used the relation,
$\eta_0\approx 2.8\times 10^{-8}\Omega_{{\rm b}} h^2$.

Carrying through the same calculation for braneworld PBHs we find 
\be
\alpha_{\rm{evap}}<3\times 10^{-10}\,
\frac{\Omega_{{\rm b}} h^2}{\mathcal{F}}
\left(\frac{g_{\rm{cos}}}{\tilde{g}^2}\right)^{1/4}
\left(\frac{M(t_{{\rm c}})}{M_4}\right)^{1/2}.
\label{alphaeHeconstraint}
\ee
Substituting the factors $\tilde{g}=0.1,\, g_{{\rm brane}}=106.75, 
\, g_{\rm{cos}}=10.75,\,\mathcal{F}=0.2$ and 
$\Omega_{{\rm b}} h^2\approx 0.02$, the observational
constraints finally are written as \cite{Zeldovichetal77}
\be
\alpha_{\rm{evap}}<1.1 \times 10^{-10}\,
\left(\frac{M_{{\rm i}}}{M_4}\right)^{1/2}\equiv {L^{{\rm ST}}_{{\rm evap}}}
\label{stcon}
\ee
and 
\be
\alpha_{\rm{evap}}<1.7 \times 10^{-10}\,
\left(\frac{M(t_{{\rm c}})}{M_4}\right)^{1/2}\equiv {L^{{\rm HE}}_{{\rm evap}}}.
\label{HEcon}
\ee

\begin{figure}[t]
\includegraphics[width=\linewidth]{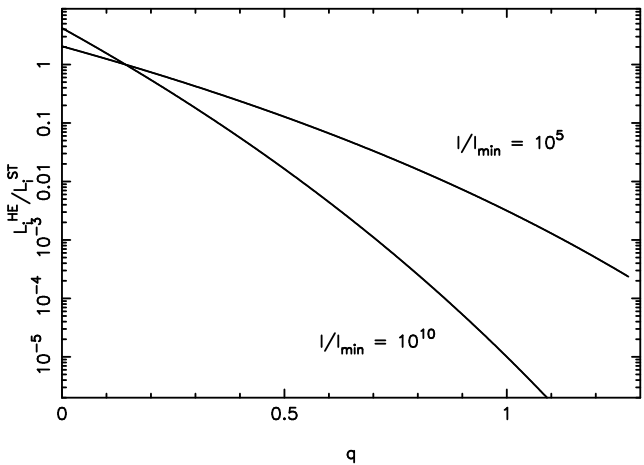}\\
\caption[limrat2]{\label{limrat2} The behaviour of Eq.~(\ref{compareinit}) as a 
function of the accretion parameter $q$ for two choices of the ratio $l/l_{{\rm 
min}}$, as appropriate to the Helium constraint. This is to be compared with 
Fig.~\ref{limrat}, as the imposed limits at evaporation in the high energy and 
standard treatments, 
$L^{{\rm HE}}_{{\rm evap}}$ and $L^{{\rm ST}}_{{\rm evap}}$,
are no longer equal.
Their ratio is instead given by 
Eq.~(\ref{Helimrat}). In the case of the Helium constraint, the limits are more 
strongly tightened in the high-energy regime than for other constraints. Note 
that for the highest values of $q$, $l$ may not be allowed to be quite as large 
as $l_{{\rm max}}$.}
\end{figure}

Making contact with Section \ref{sec:formalism}, the ratio of the
upper limits can be expressed as 
\be \frac{{L^{{\rm HE}}_{{\rm evap}}}}{{L^{{\rm ST}}_{{\rm evap}}}}\approx
\L\frac{l}{l_{{\rm min}}}\R^{-1/4}, \label{Helimrat} \ee
with $l_{{\rm min}}/l_4\approx 10^{14}-10^{15}$ for PBHs evaporating
during nucleosynthesis.
By assumption $l>l_{{\rm min}}$, hence the constraint will be
tighter in the braneworld case. This can be understood as follows: The
change in helium abundance due to evaporating PBHs is proportional to
the emitted number density, see Eq.~(\ref{deltannb}). But for a given
mass fraction $\alpha_{{\rm evap}}$, this number density is increased
for braneworld PBHs, as alluded to before. Fig.~\ref{limrat2} shows how the 
constraints are modified as compared to the standard scenario.

Using the expressions of Section~\ref{sec:formalism}, the bounds
imposed at nucleosynthesis can be converted into bounds on
the initial PBH mass fractions $\alpha_{{\rm i}}$. For standard
cosmology this gives \cite{Zeldovichetal77}
\be \alpha_{{\rm i}} < 3 \times 10^{-18}\;\L\frac{M_{{\rm
i}}}{10^9\,{\rm g}}\R^{-1/2},\ee
for PBH masses in the range $10^9 {\rm g} < M_{{\rm i}} < 10^{10} {\rm g}$.
In the braneworld scenario we obtain
\be \alpha_{{\rm i}} < 4.79 \times 10^{-11}\; \lnodim^{(1-7q)/4(4-q)}
\tLnodim^{3(q-1)/4(4-q)}, \label{he-init}\ee
with $t_{{\rm evap}}$ ranging from $1$s to $400$s.
 
For the limiting case where $l=l_{{\rm max}}=10^{31} l_4$,
neglecting accretion ($q=0$), and expressing the bound in terms of the
initial PBH mass $M_{{\rm i}}$ this gives 
\be \alpha_{{\rm i}} < 3 \times 10^{-17}\;\L\frac{M_{{\rm
i}}}{10\,{\rm g}}\R^{-3/8},\ee
with $M_{{\rm i}}$ in the range $10 {\rm g}<M_{{\rm i}}<200 {\rm g}$.
 
As remarked in Section \ref{sec:key}, for extreme values of the AdS radius
combined with a highly efficient accretion process, PBHs that are to
evaporate during the nucleosynthesis era would have been too light to
be consistent with the lower mass limit imposed by inflation. We
therefore end with an example where accretion is efficient
($q=4/\pi$), but take a more moderate value for the AdS radius,
$l=10^{25}\, l_4$. The initial mass fractions are then constrained as  
\be \alpha_{{\rm i}} < 6 \times 10^{-26}\;\L\frac{M_{{\rm
i}}}{10^{-4}\,{\rm g}}\R^{0.102},\ee
for initial masses with range $10^{-4} {\rm g}<M_{{\rm i}}<10^{-2} {\rm g}$.

%%%%%%%%%%%%%%%%%%%%%%%%%%%%%%%%%%%%%%%%%%%%%%%%%%%%%%%%%%%%%%%%%%%%%%%%%
%									%
%************************DEUTERIUM CONSTRAINT***************************%
%									%
%%%%%%%%%%%%%%%%%%%%%%%%%%%%%%%%%%%%%%%%%%%%%%%%%%%%%%%%%%%%%%%%%%%%%%%%%

\subsection{Deuterium photo-disintegration constraint}

The high-energy particles emitted by evaporating PBHs
both during and after nucleosynthesis can be sufficiently
energetic to disrupt primordial nuclei.
One important reaction of this type is so-called
{\em photo-disintegration}, or in other words, the destruction
of primordial nuclei by high-energy PBH photons.
Of all the primordial nuclei
deuterium is the most susceptible to photo-disintegration, 
since it has both the highest cross-section and also the lowest threshold, 
$Q_{{\rm d}}\sim 2.25$ MeV. A detailed analysis of this effect 
was considered by Lindley~\cite{Lindley80} for the case 
of standard four-dimensional PBHs.
Here we briefly review that work and extend
the analysis to the context of the RS-II cosmology.

Evaporation products may interact with the
background up until recombination. Therefore, in order to consider the
effects of PBH 
photo-disintegrations on the light element abundances one is generally 
concerned with PBHs evaporating between the
end of nucleosynthesis at $400$ seconds and the time of recombination
at $t_{{\rm rec}}\approx 10^{12}$ seconds. 
For standard cosmology, these are PBHs with masses in the range
\be
 10^{10}\; {\rm g} \lesssim M_{{\rm i}}\lesssim 10^{13}\; {\rm g}
\label{photomass1}
\ee
and temperatures in the range
\be
10^{3}\;{\rm GeV} \gtrsim T_{{\rm BH}}\gtrsim 1\;{\rm GeV}.
\ee
In the braneworld case both ranges will be reduced. For the extreme
case when $l=l_{{\rm max}}$, the ranges become
\be
200\; {\rm g} \lesssim  M(t_{{\rm c}}) \lesssim 10^{7}\; {\rm g}
\label{photomass2}
\ee
and
\be 200\;{\rm MeV} \gtrsim T_{{\rm BH}}\gtrsim 1\;{\rm MeV}.\ee
Although high-energy PBH photons are directly
capable of causing photo-disintegration of nuclei,
such {\em direct} disintegrations are in fact extremely rare,
since the cross-sections for photo-nuclear reactions
, even for deuterium are extremely small, typically $\lesssim 10^{-6}$.
The main effect rather is indirect and comes
instead from the photo-disintegrations caused by the
very large number of lower energy photons which
are produced as a consequence of the thermalization of
the high-energy PBH photons with the background.
This thermalization process may be understood as follows.
The high-energy photons emitted by the PBHs interact
with the background via two main processes, namely via
Compton scattering off the background electrons and via
electron-positron pair-production off the nuclei. The energetic
electrons and positrons produced in these processes then
in turn subsequently predominantly lose energy via
inverse Compton scattering off the
background photons. These photons then again
Compton scatter and pair-produce and so on.
In this way a single high-energy PBH photon
gives rise to a `cascade' of photons, electrons and
positrons of increasing number and decreasing energy.

The set of photons in each cascade, with energies $\{E_i\}$,
above the threshold $Q_{{\rm d}}$ will form a finite set, and although
these photons will all predominantly interact via Compton
scattering and pair-production processes, nevertheless will have a small
probability $P(E_i)$ of destroying a deuteron.
This probability is given by the ratio $l_\gamma/l_{{\rm d}}$ of the mean
free path of the deuteron to the photon.
If one ignores the tiny fractions of $^3\rm{He}$, $\rm{Li}$ and
$\rm{D}$, the mean free path of a photon of energy $E$
in the background is approximately \cite{Lindley80}
\be
l_\gamma^{-1}\approx n_{{\rm e}} \sigma_{{\rm c}} +(n_H +4n_{He})\sigma_{{\rm 
pp}},
\ee
where $\sigma_{{\rm c}}$ and $\sigma_{{\rm pp}}$ are respectively
the Compton-scattering and pair-production cross-sections.
Since charge neutrality demands that $n_{{\rm p}}=n_{{\rm e}}$, this may be
approximated by $l_\gamma^{-1}\approx n_{{\rm p}} \sigma_{{\rm t}}$, where
the `total cross-section' $\sigma_{{\rm t}} \approx 
\sigma_{{\rm c}}+\sigma_{{\rm pp}}$.
Hence, the probability of destroying a deuteron is given by \cite{Lindley80}
\be
P(E_{{\rm i}})=\frac{n_{{\rm d}}\sigma_{{\rm d}}(E_{{\rm i}})}{n_{{\rm 
p}}\sigma_{{\rm t}}(E_{{\rm i}})},
\ee
where $n_{{\rm d}}$ and $n_{{\rm p}}$ are the number densities of the
deuterons and
protons, and $\sigma_{{\rm d}}$ is the $d(\gamma,n)p$ cross-section.
Hence, summing over all the photons in a single cascade gives
the number of deuterons destroyed, i.e.
\be
dN_{{\rm d}}= \sum_i P(E_i).
\ee
In reality, however, each cascade is different. In order  
to take account of this variance one may introduce
an average distribution $N(E', E)$, so that the average number of
deuterons destroyed by a high-energy PBH photon with initial
energy $E$ is then given by \cite{Lindley80}
\be
dN_{{\rm d}}=-\int_{Q_{{\rm d}}}^{E} N(E',E)\,P(E') \, dE'.
\ee
In practice the form of $N(E',E)$ can be determined by numerically
modelling many cascades. Adopting such an approach,
Lindley was able to empirically estimate\footnote{Lindley also assumed that
$n_{{\rm d}}/n_{{\rm e}}$ did not change much over the thermalization timescale, 
which is
consistent with the underlying presumption that $n_{{\rm d}}$ could not have 
changed
too much.} the above integral to be of the form \cite{Lindley80}
\be
dN_{{\rm d}}=-\frac{n_{{\rm d}}}{n_{{\rm p}}}\,\beta \L\frac{E}{E_*}\R.
\label{1cascade}
\ee
Here $E_*$ and $\beta$ are constants, with ${E_*}\simeq {10^{-1} \,\rm{GeV}}$
and $\beta \sim 1$.

Thus far we have focussed only on the effect of
photo-disintegrations caused by
high-energy photons of a single energy $E$.
However, PBHs will emit photons with a
spectrum of energies. Let us consider a comoving volume $V$, containing a 
population of PBHs with total mass $M(t)$. In a time $dt$ it will evaporate
a fraction $f_{\gamma}\,dM$ of its mass into photons with a spectrum of 
energies $\nu(E)$, such that 
\be
\int E\, \nu(E) dE=-f_{\gamma} dM.
\label{fdm}
\ee
The number of deuterons in the volume $V$ destroyed in this time 
will therefore be given by integrating over
the spectrum of cascades that these evaporating photons will give rise to, i.e.
\be
dN_{{\rm d}}=-\int\int_{Q_{{\rm d}}}^{E} N(E',E)P(E')\, dE' \nu(E)\, dE.
\ee
Using Eqs.~(\ref{1cascade}) and (\ref{fdm}) we
can easily integrate this last expression to give \cite{Lindley80}
\be
\frac{N_{{\rm d}}(t_2)}{N_{{\rm d}}(t_1)}
=\frac{X_{{\rm d}}(t_2)}{X_{{\rm d}}(t_1)}
=\exp\L-\frac{\Delta M f_{\gamma} \beta}{N_{{\rm p}} E_*}\R,
\label{Nd}
\ee
where $\Delta M=M(t_1)-M(t_2)>0$ is the PBH mass evaporated
between times $t_1$
and $t_2$, $N_{{\rm d}}(t_{{\rm i}})$ is the deuteron number contained
in $V$ at time $t_{{\rm i}}$ and $X_{{\rm d}}=n_{{\rm d}}/n_{{\rm b}}$ is the 
deuteron baryon fraction. In principle, Eq.~(\ref{Nd}) should
strictly apply when PBH evaporation is the {\em only}
process responsible for changing $N_{{\rm d}}$, and should therefore
apply to the time interval
between the end of nucleosynthesis and recombination,
i.e.~from $\sim 400$ seconds until $\sim 10^{12}$ 
seconds.\footnote{In fact the range of validity can be extended 
back to the beginning of nucleosynthesis \cite{Lindley80}, but we shall not consider 
this here.}

Between the times $t_1$ and $t_2$ deuterons are destroyed, so obviously $X_{{\rm 
d}}(t_2)<X_{{\rm d}}(t_1)$. In addition, in 
order to be consistent with observational limits on the $\rm{D}$
abundance, the depletion of $X_{{\rm d}}$ cannot be too large. Hence, we
need to demand that $X_{{\rm d}}(t_2)>e^{-\epsilon}X_{{\rm d}}(t_1)$ say,  
where 
$\epsilon>0$
is some number of order unity to be constrained
by observation.\footnote{Here Lindley chooses a rough bound of $\epsilon\sim 1$,
corresponding a decrease in $X_{{\rm d}}$ of one $e$-fold over the period 
$t_2-t_1$.} With this requirement we therefore have the condition
\be
0< \frac{\Delta M f_{\gamma}\beta}{N_{{\rm p}} E_*}\lesssim \epsilon.
\ee
Taking $N_{{\rm p}} \approx M_{{\rm b}}/m_{{\rm p}}$, where $M_{{\rm b}}$ is 
the baryonic mass and $m_{{\rm p}}$ the proton mass, then \cite{Lindley80}
\be
\frac{\Delta M}{M_{{\rm b}}} \lesssim \frac{\epsilon }{f_{\gamma} 
\beta}\frac{E_*}{m_{{\rm p}}}.
\label{constraint1}
\ee

To translate Eq.~(\ref{constraint1}) into a bound on some mass fraction
of PBHs, we need to make an assumption about the PBH mass spectrum. It
is possible PBHs only form in a narrow mass range and equivalently only
evaporate at a specific era. Then $t_1$ and $t_2$ are simply taken to
contain that era and Eq.~(\ref{constraint1}) is effectively a
constraint on the total PBH mass fraction of the model. On the other
hand, PBHs could exhibit an extended, smoothly varying spectrum. In
this case the total mass evaporated is dominated by the low mass end
of the spectrum, i.e by those PBHs evaporating earliest. We may take
$t_1$ and $t_2$ to be the end of nucleosynthesis and the onset of
recombination respectively, and estimate $\Delta M$ by the PBH mass
evaporated shortly after nucleosynthesis. It is therefore usually
justified to take 
\be \frac{\Delta M}{M_{{\rm b}}}\approx \left[\frac{\rho_{{\rm pbh}}}
{\rho_{{\rm b}}}\right]_{\tevap},\ee
with $\tevap$ some time after nucleosynthesis, either when a narrow
mass range of PBHs evaporates, or straight after nucleosynthesis for
an extended mass spectrum. Eq.~(\ref{constraint1}) is then equivalent to
\be
\alpha_{\rm{evap}}
\lesssim
\left[\frac{\rho_{{\rm b}}}{\rho_{{\rm rad}}}\right]_{t_{\rm{evap}}}\, 
\frac{E_*\, \epsilon}{m_{{\rm p}}f_{\gamma}\beta}
\label{alphat}.
\ee
Furthermore, since $\rho_{{\rm b}} \propto a^{-3}$ and
$\rho_{{\rm rad}}\propto a^{-4}$, it follows that
\be
\left[\frac{\rho_{{\rm b}}}{\rho_{{\rm rad}}}\right]_{t_{\rm{evap}}} =
\frac{a(t_{\rm{evap}})}{a(t_{\rm{eq}})}\,
\left[\frac{\rho_{{\rm b}}}{\rho_{{\rm rad}}}\right]_{t_{\rm{eq}}} 
\simeq
2\frac{a(t)}{a(t_{\rm{eq}})}\,\Omega_{{\rm b}}(t_{\rm{eq}}), 
\ee
as
$\rho_{\rm{rad}} = \rho_{{\rm tot}}/2 \approx \rho_{{\rm c}}/2$ at the time of 
matter--radiation equality, $t_{\rm{eq}}\approx 8 \times 10^{54}\, t_4$.
The baryon density parameter at equality is readily related to the present one, 
as the matter density parameter at equality is given by 
$\Omega_{{\rm m}}(t_{{\rm eq}}) \approx 1/2 $: 
\be 
\Omega_{{\rm b}}(t_{\rm{eq}})= \frac{\Omega_{{\rm m}}(t_{{\rm 
eq}})}{\Omega_{{\rm m}}(t_0)}\Omega_{{\rm b}}(t_0)
\approx \frac{1}{2}\, \frac{\Omega_{{\rm b}}(t_0)}{\Omega_{{\rm m}}(t_0)} \,.
\ee

It follows that Eq.~(\ref{alphat}) may also be written as
\be
{\alpha}_{\rm{evap}}
\lesssim
\,\frac{E_*\, \epsilon}{m_{{\rm p}}f_{\gamma}\beta}\,
\L\frac{t_{\rm{evap}}}{t_{\rm{eq}}}\R^{1/2} \, 
\frac{\Omega_{{\rm b}}(t_0)}{\Omega_{{\rm m}}(t_0)}.
\label{photocon}
\ee
As a rough estimate for the fraction of the
PBH mass that decays into photons we take $f_{\gamma}=0.1$, while the
ratio of the present baryonic density to the total matter density will
be taken to be 0.1. A bound on the size of the depletion 
factor, $\epsilon$, can be given by taking the difference 
between the lowest allowed observational value and the highest 
allowed theoretical value of $\rm{D}$ from SBBNS (without PBHs), i.e.~assuming 
the validity of the quasar Lyman-$alpha$ bound 
of $\eta \ge 3.4$ as before. Following Steigman~\cite{Steigman02}, 
a cautious current observational bound on the deuterium abundance is 
$\rm{D/H}=3.0^{+1.0}_{-0.5}\times 10^{-5}$. On the other hand, a value of 
$\eta \ge 3.4$ implies a maximum SBBNS value of 
$\rm{D/H}\lesssim 7.0 \times 10^{-5}$. Thus, we may take 
$\epsilon \sim 1$, as did Lindley. Filling in all the 
parameters then finally leads to the observational constraint
\be {\alpha}_{\rm{evap}} \lesssim  3.5\times 10^{-29} \tLnodim^{1/2}
\equiv {L_{{\rm evap}}}. 
\label{deutcon}\ee
It should be noted that the constraint applies to small five-dimensional 
PBHs produced in the high-energy regime as well as to conventional 
four-dimensional PBHs. This is because the amount of deuterium
destroyed will be proportional to the total amount of emitted energy,
see Eq.~(\ref{Nd}). For a given mass fraction $\alphaevap$ the
emitted energy is identical by definition.  
 
As in the previous sections, we convert Eq.~(\ref{deutcon}) into a
constraint on an initial PBH mass fraction, and will take $\tevap=400
{\rm s}$. In standard cosmology this gives 
\be \alpha_{{\rm i}}< 10^{-21}.\ee
For braneworld black holes, taking $l=l_{{\rm max}}$ and $q=0$, we find
\be \alpha_{{\rm i}}< 10^{-16}.\ee
Finally, if $l=10^{25} l_4$ and $q=4/\pi$, the initial constraint
reads 
\be \alpha_{{\rm i}}<4 \times 10^{-26}.\ee

%%%%%%%%%%%%%%%%%%%%%%%%%%%%%%%%%%%%%%%%%%%%%%%%%%%%%%%%%%%%%%%%%%%%%%%%%%%%%%%%

% CONCLUSIONS

%%%%%%%%%%%%%%%%%%%%%%%%%%%%%%%%%%%%%%%%%%%%%%%%%%%%%%%%%%%%%%%%%%%%%%%%%%%%%%%%

\section{Conclusions}

If there were sufficiently late periods of black hole
formation in the early Universe, their presence could be noticeable in
more recent cosmological epochs such as nucleosynthesis and beyond. Observation
puts upper bounds on these effects and therefore on the allowed
abundance of PBHs, conventionally expressed as
an upper limit on the total PBH energy fraction. Using the evolution
equations for the black holes and the background cosmology, the observational
constraints can be translated into constraints on the PBH formation
rate. The initial constraints are usually the most severe for those
black holes with lifetimes comparable with the cosmic time of the
epoch at which the observational constraint is imposed. If PBHs form
from the collapse of background density perturbations, their formation
rate is related to the amplitude of the power spectrum, on scales that
enter the Hubble horizon around the time the PBHs form. Therefore, an initial
constraint implies an upper limit to the power spectrum on that scale. 

In the RS-II cosmology that we have considered, PBHs of a given lifetime
$t_{{\rm evap}}$ would have formed in the high-energy regime, provided
the AdS radius of curvature $l$ is larger than $l_{{\rm min}}\propto
(t_{{\rm evap}}/t_4)^{1/3} \,\,l_4$. Then the black hole's mass and temperature
at the onset of evaporation are reduced, and observational constraints
have to be adjusted in some cases. In addition, and in contrast
to four dimensional cosmology, the black hole is likely to grow by
accretion of the cosmic background as long as it is in the high
energy phase. Care is needed in this matter, as the growth depends
very sensitively on the efficiency of accretion.  
Such black holes will be small and effectively 5D
throughout their lifetime. 

We see that both the PBH and background evolution can be altered
compared to standard cosmology. As a consequence, the translation of
an observational limit into an upper limit on the initial PBH mass
fraction will be modified, see Eq.~(\ref{compareinit}). Most of the
effects to be constrained are simply proportional to the
total energy in PBHs, rendering the observational constraints in
standard or braneworld cosmology identical by definition.  
Eq.~(\ref{compareinit}) then allows a comparison of the strength of
the initial constraint in both scenarios. The braneworld constraint
will be the weaker if the accretion efficiency is below 50\%, 
whilst being stronger for accretion efficiencies above 50\%, 
for all values of the AdS radius $l>l_{{\rm min}}$. It should be noted that in
the latter case the initial mass of the PBHs is also much smaller than in the
standard treatment, so that the  constraint corresponds to
perturbations on smaller scales.

If PBHs were sufficiently heavy to have survived to the present day,
their mass density should not exceed that of dark matter. For PBHs
with lifetimes marginally exceeding $t_0\approx 8\times 10^{60} t_4$,
this implies a constraint on the initial PBH mass fraction
$\alpha_{{\rm i}}$. If the PBHs formed in a standard cosmological regime, the 
bound reads $\alpha_{{\rm i}} < 10^{-18}$.  If they were formed in the
high-energy regime, taking the example of $l=l_{{\rm max}}\approx
0.1 {\rm mm}$ and 100 \% accretion efficiency, the bounds strengthens
to $\alpha_{{\rm i}} < 10^{-23}$.

Primordial black holes evaporating between photon decoupling and
the present age leave behind a spectrum that peaks at a temperature of
order the black hole temperature at the onset of evaporation of PBHs
with $t_{{\rm evap}}\approx t_0$. Focussing on the photon component, it is
required that its density is less than the diffuse cosmic background
at comparable temperatures. In standard cosmology the peak temperature
lies around $100$ MeV, while for $l=l_{{\rm max}}$ it can be as low as
$200$ keV. In all cases, the photon spectrum constraint implies bounds
of roughly 9 orders of magnitude stronger than the dark matter
constraint.  

If evaporation products are released around the Sunayev--Zel'dovich
time $t_{{\rm SZ}}=10^{-10}t_0$, they will fail to fully thermalize with the
background radiation. This time, however, is sufficiently early in
order for the excess energy to distort the background blackbody
spectrum. Limits on the allowed distortion of the CMB spectrum then
imply limits on PBH mass fractions. In standard cosmology one obtains
$\alpha_{{\rm i}} < 10^{-21}$, while the extreme braneworld case
($l=l_{{\max}}$, 100 \% accretion) results in  $\alpha_{{\rm i}} < 10^{-28}$.

If there was a population of PBHs evaporating during or after the era
of nucleosynthesis ($\sim 1- 400{\rm s}$), this would have led to
changes in the final light element abundances. As a first example, we
note that the abundances depend on the neutron-to-proton ratio $n/p$ at the
onset of nucleosynthesis. A standard calculation predicts $n/p\approx
1/7$. However, due to PBH evaporation products an approximately equal
amount of (anti-) protons and neutrons is injected into
the background, increasing the neutron-to-proton ratio. This in turn
implies an increase in the helium-four mass fraction $Y_{{\rm p}}$. Comparing
the observed value with standard theoretical predictions then leads to
a PBH constraint. Furthermore, the increase in $Y_{{\rm p}}$ is proportional
to the number density of evaporation products. Recall that a PBH of
given lifetime has a reduced temperature in the braneworld
scenario. Thus, the associated constraint on the mass fraction 
$\alpha_{{\rm evap}}$ of PBHs at evaporation is strengthened in the
braneworld case, contrary to what might naively be expected. The
resulting initial constraint reads $\alpha_{{\rm i}} < 10^{-18}$ in
standard cosmology. 

For the braneworld case, the following must
be borne in mind: If there was a period of inflation, an upper
limit on its energy scale is imposed by the amount of produced
gravitational waves. The mass of subsequently formed PBHs is then
bounded from below, in turn implying a lower limit to the PBH's
lifetime. Its strength grows with the AdS radius and accretion
efficiency, and for extreme values PBH evaporation could not have been
effective until long after nucleosynthesis. We have chosen $l=10^{25}
l_4$ and
$100$ \% accretion efficiency in the initial constraint imposed by helium-four
production (Eq. (\ref{he-init})), to obtain $\alpha_{{\rm i}} < 10^{-26}$.

In a second example inspired by nucleosynthesis considerations, it was
remarked that evaporation products can be sufficiently energetic to
destroy newly formed primordial nuclei. We have focussed on 
photo-disintegration and considered the change in deuterium
abundance, since this is the nucleus most susceptible to
disintegration. The amount of deuterium that would be destroyed grows
with the total amount of injected PBH energy, showing that the ensuing
bound on $\alpha_{{\rm evap}}$ does not depend on the individual PBH
temperature. For the standard initial constraint, one arrives at    
$\alpha_{{\rm i}} < 10^{-21}$. In the example where $l=10^{25} l_4$
and accretion is maximally efficient, this becomes 
$\alpha_{{\rm i}} < 10^{-26}$.

%======================================%
%<<<<<<<<<<< ACKNOWLEDGMENTS >>>>>>>>>>%
%======================================%

\begin{acknowledgments}
D.C.~was supported by PPARC and by the EU (Marie Curie)
Development Host Fellowship HPMD-CT-2001-00070, and A.R.L.~in part by the 
Leverhulme
Trust. We thank Kazushi Iwasawa, Kasunori Khori, Bernard Pagel and 
Gary Steigman for discussions. 

\end{acknowledgments}

%======================================%
%<<<<<<<<<<<< BIBLIOGRAPHY >>>>>>>>>>>>%
%======================================%

%\addcontentsline{toc}{chapter}{Bibliography}

% Useful Journal & Book reference macros

%%%%%%%%%%%%%%%%%%%%%%%%%%%%%%%%%%%%%%%%%%%%%%%%%%%%%%%%%%%%%%%%%%%%%%%%%%%%%%%%
%%%%%%
% Standard Journal ref format
%%%%%%%%%%%%%%%%%%%%%%%%%%%%%%%%%%%%%%%%%%%%%%%%%%%%%%%%%%%%%%%%%%%%%%%%%%%%%%%%
%%%%%%

%Journal with Year:  Auth, Jrnl, Vol-No,  Page, Year.
\def\jrnld#1#2#3#4#5{{#1}, {#2} {\bf #3}, {#4} ({#5}).} % dot at end.

%Journal with Year: Auth, Jrnl, Vol-No,  Page, Year.
\def\jrnl#1#2#3#4#5{{#1}, {#2} {\bf #3}, {#4} ({#5})} % no dot.

%Single Journal:    Jrnl, VolNum,  Page, Year  % No Author and no dot.
\def\jrnlna#1#2#3#4{{#1} {\bf #2}, {#3} ({#4})}

%Journal + Title:      Auth,  Title,   Jrnl,  Vol-No,  Page,  Year.
\def\jrnlT#1#2#3#4#5#6{{#1}, {\em #2}, {#3}, {\bf #4}, {#5} ({#6}).}

%%%%%%%%%%%%%%%%%%%%%%%%%%%%%%%%%%%%%%%%%%%%%%%%%%%%%%%%%%%%%%%%%%%%%%%%%%%%%%
%Journals with E-numbers
%%%%%%%%%%%%%%%%%%%%%%%%%%%%%%%%%%%%%%%%%%%%%%%%%%%%%%%%%%%%%%%%%%%%%%%%%%%%%%

%Journal + E-No:       Auth, Jrnl, VolNum,  Page,  Year, E-No.   DOT!!
\def\jrnlEd#1#2#3#4#5#6{{#1}, {#2} {\bf #3}, {#4}  ({#5}), {\tt #6}.}

%Journal + E-No:       Auth, Jrnl, VolNum,  Page,  Year, E-No.  NO DOT!!
\def\jrnlE#1#2#3#4#5#6{{#1}, {#2} {\bf #3}, {#4}  ({#5}), {\tt #6}}

%Journal + Title & E-No:  Auth,  Title,   Jrnl,  Vol-No,  Page, Year, E-No
\def\jrnlTE#1#2#3#4#5#6#7{{#1}, {\em #2}, {#3}, {\bf #4}, {#5} ({#6}), {\tt
#7}.}

%%%%%%%%%%%%%%%%%%%%%%%%%%%%%%%%%%%%%%%%%%%%%%%%%%%%%%%%%%%%%%%%%%%%%%%%%%%%%%
% Others
%%%%%%%%%%%%%%%%%%%%%%%%%%%%%%%%%%%%%%%%%%%%%%%%%%%%%%%%%%%%%%%%%%%%%%%%%%%%%%

%Double Jour(same Auth):  Auth, Jrnl, VolNum, page  Year; Jrnl, VolNum, page
%Year.
\def\jrnltwo#1#2#3#4#5#6#7#8#9{ {#1}, {#2} {\bf #3}, {#4} (#5); {#6} {\bf #7},
{#8} {(#9)}.}

% e-print/pre-print   Auth   Title    E-No  Year
\def\preprint#1#2#3#4{{#1}, {\em #2}, {\tt #3} ({#4}).}

%%%%%%%%%%%%%%%%%%%%%%%%%%%%%%%%%%%%%%%%%%%%%%%%%%%%%%%%%%%%%%%%%%%%%%%%%%%%%
% BOOKS
%%%%%%%%%%%%%%%%%%%%%%%%%%%%%%%%%%%%%%%%%%%%%%%%%%%%%%%%%%%%%%%%%%%%%%%%%%%%%

% Book ref    :   Auth,  Title,  Press  Year.
\def\Book#1#2#3#4{{#1}, {\em #2}, {#3} (#4).}

% Book reference with specific chapter or page reference in arg{#5}
%                       Auth    Title   Pub.  Page  Year
\def\Bookpage#1#2#3#4#5{{#1}, {\em #2}, {#3}, {#4} ({#5}).}

% Book reference with specific chapter or page reference in arg{#5}
%                             Auth   Title      Book       Ed    Pub  Page  Year
\def\edBookpage#1#2#3#4#5#6#7{{#1}, {\em #2} in {#3}, Ed. {#4}, {#5}, {#6}
({#7}).}

%%%%%%%%%%%%%%%%%%%%%%%%%%%%%%%%%%%%%%%%%%%%%%%%%%%%%%%%%%%%%%%%%%%%%%%%%%%%
% PROCEEDINGS
%%%%%%%%%%%%%%%%%%%%%%%%%%%%%%%%%%%%%%%%%%%%%%%%%%%%%%%%%%%%%%%%%%%%%%%%%%%%

%proceedings:           Auth,  Title,   Proceedings,   Eds,    Pub   Year  E-No
\def\Proc#1#2#3#4#5#6#7{{#1}, {\em #2}, Proc. {#3 }, {Ed. #4}, {#5} ({#6}), {\tt
#7}.}

%\baselinestretch{1}}

%%%%%%%%%%%%%%%%%%%%%%%%%%%%%%%%%%%%%%%%%%%%%%%%%%%%%%%%%%%%%%%%%%%%%%%%%%%%
% Some useful journal abbreviations
%%%%%%%%%%%%%%%%%%%%%%%%%%%%%%%%%%%%%%%%%%%%%%%%%%%%%%%%%%%%%%%%%%%%%%%%%%%%

\def\ADM{Adv. Math.}				
\def\ADS{Adv. Space Res.} 			
\def\ANP{Ann. Phys. (NY)}
\def\ARM{Arch. Math.}
\def\ASA{Astron. Astrophys.}
\def\ASJ{Astrophys. J.}
\def\ASP{Astropart. Phys.}
\def\ASS{Astrophys. Space Sci.}
\def\ASZ{Astron. Zh.}
\def\CMP{Commun. Math. Phys.}
\def\CQG{Class. Quant. Grav.}
\def\GRG{Gen. Rel. Grav.}
\def\IJM{Int. J. Mod. Phys.}
\def\JHEP{JHEP}
\def\JLM{J. London Math. Soc.}
\def\JMP{J. Math. Phys.}
\def\JOP{J. Phys.}
\def\MAA{Math. Ann.}
\def\MNR{Mon. Not. Roy. Ast. Soc.}
\def\MPL{Mod. Phys. Lett.}
\def\MAZ{Math. Zeit.}
\def\NUC{Nuovo Cimento}
\def\NIM{Nucl. Instrum. Meth.}
\def\NAT{Nature}
\def\NUP{Nucl. Phys.}
\def\PAZ{Pis'ma Astron. Zh.}
\def\PHL{Phys. Lett.}
\def\PHR{Phys. Rep.}
\def\PHR{Phys. Rev. D}
\def\PHS{Physica Scripta}
\def\PRL{Phys. Rev. Lett.}
\def\PRS{Proc. Roy. Soc. Lon.}
\def\PTR{Phil. Trans. Roy. Soc. Lon.}
\def\PTP{Prog. Theor. Phys.}
\def\PZE{Pis'ma Zh. Eksp. Teor. Fiz}
\def\RAP{Rel. Astro. Phys.}
\def\RMP{Rev. Mod. Phys.}
\def\RPP{Rep. Prog. Phys.}
\def\SAL{Sov. Astron. Lett.}
\def\SOA{Sov. Astron.}
\def\SPJ{Sov. Phys. JETP}
\def\TEN{Tensor N. S.}
\def\ZEP{Z. Phys.}
\def\ZET{Zh. Eksp. Teor. Fiz.}

% KEY FOR JOURNAL ABBREVIATIONS
% 1 word title: 1st 3 letters of title, e.g. Nature=\NAT
% 2 word title: 1st 2 of 1st word + 1st of 2nd, e.g. Astrophysical Journal=\ASJ
% 3+word title: 1st letter of 1st three words, e.g. Reviews of Modern
%Physics=\RMP

%%%%%%%%%%%%%%%%%%%%%%%%%%%%%%%%%%%%%%%%%%%%%%%%%%%%%%%%%%%%%%%%%%%%%%%%%%%%%
% Some Book Publishers
%%%%%%%%%%%%%%%%%%%%%%%%%%%%%%%%%%%%%%%%%%%%%%%%%%%%%%%%%%%%%%%%%%%%%%%%%%%%%

\def\ACP{Academic Press Inc.}
\def\ADW{Addison-Wesley Publishing Company: Redwood City, California}
\def\CUP{Cambridge University Press: Cambridge}
\def\DOV{Dover Publications, New York}
\def\ELS{Elsevier Science B. V., Amsterdam}
\def\FRE{W. H. Freeman, New York}
\def\HUP{Harvard University Press: Cambridge, Massachusetts}
\def\IOP{Institute of Physics Publishing: Bristol and Philadelphia}
\def\KLU{Kluwer: Dordrecht}
\def\MCH{McGraw-Hill Book Company: New York}
\def\PUP{Princeton University Press, Princeton: New Jersey}
\def\SVN{Springer-Verlag: New York}
\def\SVB{Springer-Verlag: Berlin-Heidelberg}
\def\UCP{University of Chicago Press: Chicago}
\def\WIL{John Wiley and Sons Ltd: Chichester}
\def\WOS{World Scientific: Singapore}

%Same key as above

%%%%%%%%%%%%%%%%%%%%%%%%%%%%%%%%%%%%%%%%%%%%%%%%%%%%%%%%%%%%%%%%%%%%%%%%

%%%%%%%%%%%%%%%%%%%%%%%%%%%%%%%%%%%%%%%%%%%%%%%%%%%%%%%%%%%%%%%%%%%%%%%%
\end{document}